\documentclass[aps,prd,amsfonts,amssymb, amsmath, showpacs, showkeys, a4paper, nofootinbib, superscriptaddress, 11pt, raggedbottom]{revtex4-2}

\usepackage{braket}
\usepackage{bm}
\usepackage{graphicx}
\usepackage{slashed}
\usepackage{subfig}
\usepackage{mathrsfs}
\usepackage{color}
\usepackage{mathtools}
\usepackage[normalem]{ulem}
\usepackage{simplewick}

\newcommand{\kket}[1]{| #1 \rangle\!\rangle}
\newcommand{\bbra}[1]{\langle\!\langle #1 |}

\begin{document}

\title{Particles in finite volumes and a toy model of decaying neutrons}

\author{Christian K\"{a}ding}
\email{christian.kaeding@tuwien.ac.at}
\affiliation{Atominstitut, Technische Universit\"at Wien, Stadionallee 2, 1020 Vienna, Austria}

\begin{abstract}
It is well-known that the momentum spectra of particles confined to finite spatial volumes deviate from the continuous spectra used for unconfined particles. In this article, we consider real scalar particles confined to finite volumes with periodic boundary conditions, such that the particles' spectra are discrete. We directly compute the density matrices describing the decay processes $\phi \to \varphi^2$ and $\phi \to \varphi\chi\nu$, and subsequently derive expressions for the decay probabilities both for confined and unconfined particles. The latter decay process is used as a rough toy model for a neutron decaying into a proton, an electron, and an anti-electron neutrino. We propose that finite volume effects can have an impact on the outcomes of experiments measuring the neutron lifetime. In addition, our findings at the toy model level suggest that taking into account possible initial correlations between neutrons and their daughter particles might be relevant as well. 
\end{abstract}


\maketitle



\section{Introduction}

In many computations, it is assumed that the energy and momentum spectra of particles are continuous. However, particles are often confined within finite volumes, for example, within a finite time interval from a particle's creation to its annihilation or within a region of space. Depending on the boundary conditions, a restriction to finite volumes renders particles' spectra different from those in the infinite volume case. For instance, a particle spatially confined to a cube with periodic boundary conditions will have a discrete spectrum depending on the dimensions of the cube. Such differences between finite and infinite volume cases can have physical consequences, for example, the Casimir effect \cite{Casimir:1948dh}, the dynamical Casimir effect \cite{Moore:1970tmc} or the Purcell effect \cite{purcell1946}. Important theoretical work on particles and fields in finite volumes has been done by L\"uscher in Refs.~\cite{Luscher:1985dn,Luscher:1986pf,Luscher:1990ux,Luscher:1991cf,Lellouch:2000pv}, while more recent discussions considering finite volume effects in a variety of research areas can be found, for example, in Refs.~\cite{Takagi1992,Bazhanov:1996aq,Meyer:2009kn,Gromov:2008gj,Polejaeva:2012ut,Kreuzer:2012sr,Briceno:2015tza,Barbado:2018qod,Romero-Lopez:2019qrt,Juarez-Aubry:2020aoo,Barbado:2021wnn,Guo2021,Blanton:2021mih,Zhao:2021ahi,Klauder:2022gke,Bajnok:2023bpj,Ai:2024taz,Alexandre:2024htk,GarciaMartin-Caro:2024qpk,Alexandre:2024gat}.

An important open problem in physics is the correct experimental determination of the mean lifetime $\tau$ of free neutrons. The neutron lifetime is not only connected
to fundamental parameters of the Standard Model \cite{Auler:2025mbm}, but also has implications for cosmology \cite{Mathews:2004kc,Chowdhury:2022ahn}; see also Ref.~\cite{Wietfeldt:2024oku}. Therefore, it is problematic that, to date, no concordance on the value of $\tau$ has been reached; see Ref.~\cite{Abele:2008zz} for a good historic overview. More precisely, there is an about $10$ s discrepancy between the two established main methods of determining the neutron lifetime: the beam method \cite{Robson1951,Nico:2004ie,Yue:2013qrc} and the ultra-cold neutron (UCN) storage method \cite{Morris2017,Pattie:2017vsj,UCNt:2021pcg,Serebrov:2017bzo,Mampe:1993an,SEREBROV200572,Pichlmaier:2010zz,Steyerl:2012zz,Arzumanov:2015tea,Ezhov:2014tna,Musedinovic:2024gms}. The former finds on average $\tau = 888.1 \pm 2.0$ s, while the results of the latter average to $\tau = 878.36 \pm 0.45$ s \cite{Wietfeldt:2024oku}. Note that there are also indirect ways of determining the neutron lifetime experimentally; see, for example, Refs.~\cite{Abele:2002wc,Mund:2012fq,Markisch:2018ndu}. Furthermore, a recent alternative beam measurement found a result closer to those of the UCN storage method experiments \cite{Fuwa:2024cdf}. Nevertheless, the neutron lifetime problem remains unresolved. While it is widely assumed that yet unknown systematic errors in the experimental methods are responsible for the discrepancy \cite{Wietfeldt:2024oku}, there have also been suggestions of new exotic physics as possible explanations \cite{Altarev:2009tg,Barducci:2018rlx,Berezhiani:2018eds,Berezhiani:2018xsx,Dubbers:2018kgh,Klopf:2019afh,Belfatto:2019swo,Giacosa:2019nbz,Tan:2023mpj,Fornal:2023wji,Dvali:2023zww,Koch:2024cfy, OKS2024102275}. Furthermore, it has been proposed that the neutron lifetime discrepancy is caused by an increase in the decay probability due to multiple elastic collisions between neutrons and between neutrons and trap walls \cite{Desai+2025}.

In this article, we make a new suggestion that might contribute to the resolution of the neutron lifetime problem. We propose that finite volume effects can actually be of experimental relevance when measuring the neutron lifetime. Since the neutron lifetime experiments typically have different confinement structures, e.g., for the beam method the neutron is essentially unconfined in at least one direction \cite{Wietfeldt:2024oku}, while for the UCN storage method used in the upcoming $\tau$SPECT experiment \cite{Auler:2023tuf,Auler:2025mbm} the neutron is confined in all spatial directions, it is possible that there naturally are noticeable differences in the decays of free neutrons depending on the experimental setup. Note that this does not require the introduction of any new physics, but rather a more careful theoretical analysis taking into account the computational differences between finite and infinite volume cases as well as the different boundary conditions of each experiment resulting from properties of the confining boundaries, e.g., how likely it is that an interaction with them leads to losses of neutron energies or even single neutrons. Similar ideas considering a possible dependence of the neutron lifetime on the experimental environment were recently discussed in Ref.~\cite{He:2021wmq}, where a measurable neutron Purcell effect has been proposed, and in Ref.~\cite{Cea:2021osz}, where it was suggested that invoking the Casimir effect for trapped ultra-cold neutrons can resolve the neutron lifetime problem. The latter computes the neutron lifetime from the usual transition amplitude approach and only considers the vacuum energy density shift, that arises in finite volumes due to the Casimir effect, as an additional contribution to the Fermi phase-space factor. In the present article, we derive the neutron lifetime from the directly computed density matrix describing the decay process and we take into account that the discrete spectrum arising in finite volumes requires us to work with sums over 3-momenta instead of integrals. In this way, our approach should be capable of capturing all possible finite volume corrections also beyond the one described in Ref.~\cite{Cea:2021osz}. Though, we will neglect finite volume corrections to masses since they can only appear at orders in perturbation theory higher than what we will consider here.

The article is structured as follows. In Sec.~\ref{sec:pre}, we will introduce a few mathematical prerequisites for the computations in finite and infinite volumes. Subsequently, in Sec.~\ref{sec:two}, we will discuss the decay of a single scalar particle $\phi$ into two copies of another scalar $\varphi$, both in infinite and finite volumes. More precisely, we will derive the density matrices describing this decay for both types of volumes and then compare the resulting expressions for the decay probabilities. This computation will allow us to get an idea of how to approximate certain integrals that we will later also encounter when discussing neutrons. For this computation and also for all later ones, we will employ methods that were developed in Refs.~\cite{Burrage:2018pyg,Burrage:2019szw,Kading:2022jjl,Kading:2022hhc}, which are in turn based on the Schwinger-Keldysh formalism \cite{Schwinger,Keldysh} and thermo field dynamics (TFD) \cite{Takahasi:1974zn,Arimitsu:1985ez,Arimitsu:1985xm, Khanna}, and have already found phenomenological applications in Refs.~\cite{Kading:2023mdk,Fahn:2024fgc,Kading:2024jqe,Burrage:2025xac}; see also Ref.~\cite{Kading:2025cwg} for an alternative introduction and discussion of these methods. Next, in Sec.~\ref{sec:neutrons}, we will introduce a rudimentary toy model for decaying neutrons. More specifically, we again consider a real scalar $\phi$ (a 'neutron') that decays into three other scalar particles $\varphi$ (a 'proton'), $\chi$ (an 'electron') and $\nu$ (an 'anti-electron neutrino'). We will consider the neutron and its decay products to be confined to a finite spatial volume with perfectly reflecting boundaries like the magnetic trap in the upcoming $\tau$SPECT experiment \cite{Auler:2023tuf,Auler:2025mbm}, and again compute the decay probability in this finite volume. From this naive computation, we will derive an extremely large predicted neutron lifetime. As a consequence, we will improve the toy model by taking into account that the anti-neutrino is essentially not confined, which requires us to treat it as having a continuous spectrum, and by considering that the neutron and its decay products will likely be correlated in Fock space. With those two improvements we will manage to predict a neutron lifetime that, for a simple toy model, is impressively close to results of real neutron lifetime experiments. We will interpret this as a hint that finite volume effects are indeed of relevance to the neutron lifetime problem. Finally, in Sec.~\ref{sec:Conclusion}, we will draw our conclusions and give an outlook on further possible improvements required for evolving the toy model into a more realistic model that can actually confirm our proposal.


\section{Scalar fields in finite and infinite volumes}
\label{sec:pre}

In this section, we will shortly introduce the mathematical prerequisites for our computations for particles and fields in finite and infinite volumes. We base our introduction on Refs.~\cite{huang2010quantum,Peterken:2024ktm}. 

In case of an infinite volume, a scalar field operator on either the $+$ or $-$ branch of a Schwinger-Keldysh closed time path \cite{Schwinger,Keldysh} can be expanded in terms of creation and annihilation operators as  
\begin{eqnarray}
\label{eq:FieldOperInf}
    \hat{\phi}^\pm_x &=& \int d\Pi^\phi_\mathbf{k}
\left[ 
\hat{a}^\pm_{\mathbf{k}} e^{\pm\mathrm{i}(\mathbf{k}\mathbf{x}-E^\phi_\mathbf{k} t)} 
+
\hat{a}^{\pm\dagger}_{\mathbf{k}} e^{\mp\mathrm{i}(\mathbf{k}\mathbf{x}-E^\phi_\mathbf{k} t)}
\right]~,
\end{eqnarray}
where 
\begin{eqnarray}
    d\Pi^\phi_\mathbf{k} &:=& \frac{d^3k }{(2\pi)^3 2 E^\phi_\mathbf{k}}~, 
\end{eqnarray}
and $E^\phi_\mathbf{k} = \sqrt{\mathbf{k}^2 + M^2}$ is the on-shell energy of a $\phi$-particle with 3-momentum $\mathbf{k}$ and mass $M$. If we instead consider a finite volume $V= L_x  L_y  L_z$ with periodic boundary conditions, Eq.~(\ref{eq:FieldOperInf}) becomes 
\begin{eqnarray}
\label{eq:FieldOperFin}
    \hat{\phi}^\pm_x &=& \frac{1}{V}\sum_\mathbf{k}\frac{1}{  2 E_\mathbf{k}^\phi} 
\left[ 
\hat{a}^\pm_{\mathbf{k}} e^{\pm\mathrm{i}(\mathbf{k}\mathbf{x}-E_\mathbf{k}^\phi t)} 
+
\hat{a}^{\pm\dagger}_{\mathbf{k}} e^{\mp\mathrm{i}(\mathbf{k}\mathbf{x}-E_\mathbf{k}^\phi t)}
\right]~,
\end{eqnarray}
where we now have a discrete momentum spectrum with vector components $k_i = \frac{2\pi}{L_i}n_i$. The integers $n_i$ are components of a vector $\mathbf{n} \in \mathbb{Z}^3$. In the limit $V \to \infty$, we recover Eq.~(\ref{eq:FieldOperInf}) from Eq.~(\ref{eq:FieldOperFin}) since we make the replacement
\begin{eqnarray}
     \frac{1}{V}\sum_\mathbf{k} \to \int \frac{d^3k }{(2\pi)^3}~.
\end{eqnarray}
Furthermore, for $V \to \infty$, we have
\begin{eqnarray}
     V \delta_{\mathbf{k},\mathbf{k'}} \to (2\pi)^3 \delta^{(3)}(\mathbf{k}-\mathbf{k'})
    ~,
\end{eqnarray}
such that 
\begin{eqnarray}
\label{eq:commutator}
    \left[\hat{a}^\pm_{\mathbf{k}}, \hat{a}^{\pm\dagger}_{\mathbf{k'}}\right] &=&  2 E^\phi_\mathbf{k} V \delta_{\mathbf{k},\mathbf{k'}} 
\end{eqnarray}
in finite volumes. Finally, another important relation for us is
\begin{eqnarray}
\label{eq:xIntdisc}
    \int_{-L_i/2}^{L_i/2} d^3x e^{\mathrm{i}(\mathbf{k} - \mathbf{k}')\mathbf{x}} &=&  
    V \delta_{\mathbf{k},\mathbf{k'}}~.
\end{eqnarray}


\section{Two scalar fields}
\label{sec:two}

Before moving on to the neutron decay toy model, we will discuss a simpler example in order to illustrate differences between computations in finite and infinite volumes and to learn more about the approximations that we will be applying throughout this article. We consider a real scalar field $\phi$ with mass $M$ that interacts
via 
\begin{eqnarray}
\label{eq:Action12}
S_{\text{int}}[\phi;\varphi] 
    &=&
    \int_{x\in\Omega_{t,V}} 
    \left[ - \alpha\mathcal{M} \phi\varphi^2   \right]
\end{eqnarray}
with another real scalar field $\varphi$ that has a mass $m_\varphi$. Apart from this interaction, the two scalars are free and have actions
\begin{eqnarray}
\label{eq:phivarphiac}
    S_\phi[\phi] &=& \int_x \left[ -\frac{1}{2}(\partial \phi)^2 -\frac{1}{2} M^2\phi^2 \right]~,~~~
      S_\varphi[\varphi] = \int_x \left[ -\frac{1}{2}(\partial \varphi)^2 -\frac{1}{2} m_\varphi^2\varphi^2 \right]~,
\end{eqnarray}
such that the total action is given by $S[\phi;\varphi] =S_\phi[\phi]  + S_\varphi[\varphi] + S_{\text{int}}[\phi;\varphi] $.
Here, we have introduced the notation
\begin{eqnarray}
    \int_x &:=& \int d^4x~.
\end{eqnarray}
The mass scale $\mathcal{M}$ is left undetermined for our discussion, $\alpha \ll 1$ is a dimensionless coupling constant, and we define the set
${\Omega_{t,V} := [0,t]\times [-L_x/2,L_x/2] \times [-L_y/2,L_y/2]\times[-L_z/2,L_z/2]}$, which includes the case $L_i \to \infty$.

We will use Ref.~\cite{Kading:2022hhc} and lend tools from TFD \cite{Takahasi:1974zn,Arimitsu:1985ez,Arimitsu:1985xm, Khanna} in order to find the density matrix elements describing the decay of a single $\phi$-particle into two $\varphi$-particles at second order in $\alpha$ first in an infinite volume and then for a finite volume $V$. Subsequently, we will derive expressions for the decay probabilities in both cases. Comparing the two results with each other, we will notice differences between them. Finally, relating our results to those in Ref.~\cite{Lellouch:2000pv} will tell us more about the approximations that we will have employed.


\subsection{Density matrix elements}
\label{sec:DME12}

\begin{figure}[htbp]
\begin{center}
\includegraphics[scale=0.6]{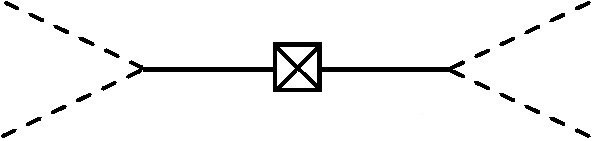}
\caption{Taken from Ref.~\cite{Kading:2022hhc}; the crossed box represents the single $\phi$-particle density matrix elements at the initial time $0$, a solid line is a $\phi$-propagator, and a dotted line stands for a $\varphi$-propagator. To the left and to the right of the crossed box, time evolves from $0$ to the final time $t$. More precisely, the left-hand side of this diagram shows the evolution $\bra{\mathbf{p}+\mathbf{k};;0} \to  \bra{;\mathbf{p}, \mathbf{k};t}$ and the right-hand side depicts $\ket{\mathbf{p}'+\mathbf{k}';;0} \to \ket{;\mathbf{p}', \mathbf{k}';t}$.}
\label{fig:1to2}
\end{center}
\end{figure}

Ref.~\cite{Kading:2022hhc} has also dealt with the example considered here and already provides a result for the infinite volume density matrix elements describing the decay of a single $\phi$ into two copies of $\varphi$:
\begin{eqnarray}
\label{eq:oldresult}
\rho^\infty_{0,2;0,2}(;\mathbf{p}, \mathbf{k}|;\mathbf{p}', \mathbf{k}'|t)
&\approx&
\frac{\alpha^2 \mathcal{M}^2}{4}
\frac{\rho^\infty_{1,0;1,0}(\mathbf{p}+\mathbf{k};|\mathbf{p}'+\mathbf{k}';|0)}{E^\phi_{\mathbf{p}+\mathbf{k}}E^\phi_{\mathbf{p}'+\mathbf{k}'}(E^\phi_{\mathbf{p}+\mathbf{k}} - E^\varphi_{\mathbf{p}} -E^\varphi_{\mathbf{k}})(E^\phi_{\mathbf{p}'+\mathbf{k}'} - E^\varphi_{\mathbf{p}'} -E^\varphi_{\mathbf{k}'})}
\nonumber
\\
&&
~~~~~~~~~~
\times
\bigg[  
e^{-\mathrm{i}(E^\varphi_{\mathbf{p}} + E^\varphi_{\mathbf{k}})t}
-
e^{-\mathrm{i}E^\phi_{\mathbf{p}+\mathbf{k}}t}
\bigg]
\bigg[  
e^{\mathrm{i}(E^\varphi_{\mathbf{p}'} + E^\varphi_{\mathbf{k}'})t}
-
e^{\mathrm{i}E^\phi_{\mathbf{p}'+\mathbf{k}'}t}
\bigg]
~;
\end{eqnarray}
see App.~\ref{app:SK} for more details on how to obtain this result. Note that we have only considered the connected diagram in Fig.~\ref{fig:1to2}, but have dropped all disconnected diagrams. We have used $\approx$ in order to illustrate that this result is only valid at $\mathcal{O}(\alpha^2)$. The infinite volume density matrix elements $\rho^\infty_{0,2;0,2}$ and $\rho^\infty_{1,0;1,0}$ represent the two $\varphi$-particle states and the single $\phi$-particle states, respectively, and are obtained by projecting the total density operator into the respective subspaces in the Fock basis:
\begin{eqnarray}
   \rho^\infty_{0,2;0,2}(;\mathbf{p}, \mathbf{k}|;\mathbf{p}', \mathbf{k}'|t) 
   &=&
   \bra{;\mathbf{p}, \mathbf{k};t} \hat{\rho}(t)\ket{;\mathbf{p}', \mathbf{k}';t}
   ~,
   \\
   \rho^\infty_{1,0;1,0}(\mathbf{p}+\mathbf{k};|\mathbf{p}'+\mathbf{k}';|0)
   &=&
   \bra{\mathbf{p}+\mathbf{k};;0} \hat{\rho}(0)\ket{\mathbf{p}'+\mathbf{k}';;0}~.
\end{eqnarray}
Since there is no result for the finite volume computation of these density matrix elements in the literature, we need to derive it ourself. For this, we follow Ref.~\cite{Kading:2022hhc} and lend tools from TFD, such that we can start from 
\begin{eqnarray}
\label{eq:Startdisc12}
\rho^V_{0,2;0,2}(;\mathbf{p}, \mathbf{k}|;\mathbf{p}', \mathbf{k}'|t)
&=&
\mathrm{Tr}\ket{;\mathbf{p}',\mathbf{k}'}\bra{;\mathbf{p},\mathbf{k}} \hat{\rho}(t)
\nonumber
\\
&=&
\langle\langle 1|(\ket{;\mathbf{p}',\mathbf{k}'}\bra{;\mathbf{p},\mathbf{k}}\otimes \hat{\mathbb{I}})\hat{\rho}^+(t) |1\rangle\rangle
\nonumber
\\
&\approx&
\langle\langle 1|(\ket{;\mathbf{p}',\mathbf{k}'}\bra{;\mathbf{p},\mathbf{k}}\otimes \hat{\mathbb{I}})\sum\limits_{a,b=\pm} ab \hat{S}^a_{\text{int}}(t) \hat{S}^b_{\text{int}}(t) \hat{\rho}^+(0) |1\rangle\rangle
\nonumber
\\
&\approx&
- \frac{\alpha^2}{8}\mathcal{M}^2
\langle\langle ;\mathbf{p}_+,\mathbf{k}_+, \mathbf{p}'_-,\mathbf{k}'_-;t|
 \sum\limits_{a,b=\pm} ab \int_{zz'} \hat{\phi}_z^a\hat{\phi}_{z'}^b(\hat{\varphi}_z^a)^2(\hat{\varphi}_{z'}^b)^2
\hat{\rho}^+(0) |1\rangle\rangle~,~~~~~
\end{eqnarray}
where we are now working in the usual TFD doubled Hilbert space $\widehat{\mathcal{H}} := \mathcal{H}^+ \otimes \mathcal{H}^-$ with operators 
\begin{eqnarray}
\hat{\mathcal{O}}^+ &:=& \hat{\mathcal{O}} \otimes \hat{\mathbb{I}}~~,~~
\hat{\mathcal{O}}^- := \hat{\mathbb{I}} \otimes \hat{\mathcal{O}}^\mathcal{T}~,
\end{eqnarray}
for which $\mathcal{T}$ indicates time-reversal. 
The operator $\hat{S}_{\text{int}}(t)$ corresponds to the interaction action in Eq.~(\ref{eq:Action12}). 
For the $\phi$- and $\varphi$-fields we introduce creation and annihilation operators $\hat{a}^\dagger$, $\hat{a}$ and $\hat{b}^\dagger$, $\hat{b}$, respectively. The creators act on the TFD vacuum state
\begin{eqnarray}
    \kket{0} &:=& \ket{;} \otimes \ket{;}
\end{eqnarray}
as
\begin{eqnarray}
\hat{a}^{+\dagger}_\mathbf{k} \kket{0} &=& \ket{\mathbf{k};} \otimes \ket{;} \,=:\, \kket{\mathbf{k}_+;}
~,~~~
\hat{a}^{-\dagger}_\mathbf{k} \kket{0} \,=\, \ket{;}\otimes \ket{\mathbf{k};} \,=:\, \kket{\mathbf{k}_-;}~,
\nonumber
\\
\hat{b}^{+\dagger}_\mathbf{k} \kket{0} &=& \ket{;\mathbf{k}} \otimes \ket{;} \,=:\, \kket{;\mathbf{k}_+}
~,~~~
\hat{b}^{-\dagger}_\mathbf{k} \kket{0} \,=\, \ket{;} \otimes \ket{;\mathbf{k}} \,=:\, \kket{;\mathbf{k}_-}~,
\end{eqnarray}
while the annihilators give
\begin{eqnarray}
\hat{a}^{\pm}_\mathbf{k} \kket{\mathbf{p}_+,\mathbf{p}_-;} \,=\, 2 E^\phi_\mathbf{k} V \delta_{\mathbf{p},\mathbf{k}} \kket{\mathbf{p}_\mp;}
~,~~~
\hat{b}^{\pm}_\mathbf{k} \kket{;\mathbf{p}_+,\mathbf{p}_-} \,=\,  2 E^\varphi_\mathbf{k}V \delta_{\mathbf{p},\mathbf{k}}\kket{;\mathbf{p}_\mp}
~.~~~~
\end{eqnarray}
In addition, we define a state
\begin{eqnarray}
\kket{1} &:=& \kket{0} +  \sum_{\mathbf{k}}\frac{1}{2 E^\phi_\mathbf{k}} \kket{\mathbf{k}_+,\mathbf{k}_-;}
+  \sum_{\mathbf{l}}\frac{1}{2 E^\varphi_\mathbf{l}} \kket{;\mathbf{l}_+,\mathbf{l}_-}
+  \sum_{\mathbf{k},\mathbf{l}}\frac{1}{4 E^\phi_\mathbf{k}E^\varphi_\mathbf{l}} \kket{\mathbf{k}_+,\mathbf{k}_-;\mathbf{l}_+,\mathbf{l}_-} 
\nonumber
\\
&&
~~~~~
+ \frac{1}{2!}\sum_{\mathbf{k}\mathbf{k'}}
\frac{1}{4 E^\phi_\mathbf{k}E^\phi_\mathbf{k'}}
\kket{\mathbf{k}_+,\mathbf{k'}_+,\mathbf{k}_-,\mathbf{k'}_-;}+...~~~,
\end{eqnarray}
which is picture-independent \cite{Kading:2022jjl} and allows us to take traces over operators within TFD; see, for example, Ref.~\cite{Kading:2025cwg} for more details on this formalism. 

Using Eqs.~(\ref{eq:commutator}) and (\ref{eq:xIntdisc}), Eq.~(\ref{eq:Startdisc12}) leads us to
\begin{eqnarray}
\label{eq:disc12deres}
\rho^V_{0,2;0,2}(;\mathbf{p}, \mathbf{k}|;\mathbf{p}', \mathbf{k}'|t)
&\approx&
 \frac{\alpha^2}{4}\mathcal{M}^2
\langle\langle ;\mathbf{p}_+,\mathbf{k}_+, \mathbf{p}'_-,\mathbf{k}'_-;t|
  \int_{zz'} 
  \frac{1}{V^8}\sum_{\mathbf{q}\mathbf{q}'\mathbf{l}\mathbf{r}\mathbf{s}\mathbf{u}\mathbf{v}\mathbf{w}}\frac{1}{  2^8 E^\phi_{\mathbf{q}} E^\phi_{\mathbf{q}'} E^\phi_{\mathbf{l}} E^\phi_{\mathbf{r}} E^\varphi_{\mathbf{s}} E^\varphi_{\mathbf{u}}
   E^\varphi_{\mathbf{v}} E^\varphi_{\mathbf{w}}} 
      \nonumber
 \\
 &&
 \times
\hat{a}^+_{\mathbf{l}} e^{\mathrm{i}({\mathbf{l}}\mathbf{z}-E^\phi_{\mathbf{l}} z^{0})}
\hat{a}^-_{\mathbf{r}} e^{-\mathrm{i}(\mathbf{r}\mathbf{z}'-E^\phi_\mathbf{r} z^{0\prime})} 
\hat{b}^{+\dagger}_{\mathbf{s}}\hat{b}^{+\dagger}_{\mathbf{u}} e^{-\mathrm{i}((\mathbf{s}+\mathbf{u})\mathbf{z}-(E^\varphi_\mathbf{s} + E^\varphi_\mathbf{u}) z^{0})}
\nonumber
 \\
 &&
 \times
\hat{b}^{-\dagger}_{\mathbf{v}}\hat{b}^{-\dagger}_{\mathbf{w}} e^{\mathrm{i}((\mathbf{v}+\mathbf{w})\mathbf{z}'-(E^\varphi_\mathbf{v} + E^\varphi_\mathbf{w}) z^{0\prime})}
\rho^V_{1,0;1,0}(\mathbf{q};|\mathbf{q}';|0) |\mathbf{q}_+,\mathbf{q}'_-;\rangle\rangle
\nonumber
\\
&\approx&
 \frac{\alpha^2 \mathcal{M}^2 \rho^V_{1,0;1,0}(\mathbf{p}+\mathbf{k} ;|\mathbf{p}'+\mathbf{k}';|0) }{4 E^\phi_{\mathbf{p}+\mathbf{k}}E^\phi_{\mathbf{p}'+\mathbf{k}'}(E^\phi_{\mathbf{p}+\mathbf{k}}-E^\varphi_\mathbf{p} - E^\varphi_\mathbf{k})(E^\phi_{\mathbf{p}'+\mathbf{k}'}-E^\varphi_{\mathbf{p}'} - E^\varphi_{\mathbf{k}'})}
 \nonumber
\\
&&
\times
\bigg[  
e^{-\mathrm{i}(E^\varphi_{\mathbf{p}} + E^\varphi_{\mathbf{k}})t}
-
e^{-\mathrm{i}E^\phi_{\mathbf{p}+\mathbf{k}}t}
\bigg]
\bigg[  
e^{\mathrm{i}(E^\varphi_{\mathbf{p}'} + E^\varphi_{\mathbf{k}'})t}
-
e^{\mathrm{i}E^\phi_{\mathbf{p}'+\mathbf{k}'}t}
\bigg]
~.
\end{eqnarray} 
A comparison with Eq.~(\ref{eq:oldresult}) shows us that the density matrix elements are essentially the same for the infinite and finite volume cases, apart from the initial density matrix elements used. Why this happens can easily be understood. For this, we have to remember that the relevant difference between the computation in App.~\ref{app:SK}, which leads to the result in Eq.~(\ref{eq:oldresult}), and the computation, which gives us Eq.~(\ref{eq:disc12deres}), is the fact that for the former we have to integrate over 3-momenta, but for the latter we sum over them. All other aspects of the two computations are essentially the same. In App.~\ref{app:SK}, we have seen that, for our tree-level calculation, all 3-momentum integrals are over Dirac delta functions that result from integrations over the spatial coordinates. For the computation in a finite volume, the sums over 3-momenta are evaluated in the same way by using Kronecker deltas obtained from expressions like the one in Eq.~(\ref{eq:xIntdisc}). Effectively, this leads in both cases to the same results. Though, if we go beyond the tree-level and consider mass correcting diagrams with internal loops, then there are not sufficiently many integrals over spatial coordinates that can give us delta functions or Kronecker deltas in order to evaluate all 3-momentum integrals or sums, respectively. In such a case, we would actually find a real difference in the results for computations in infinite and finite volumes. However, in the model considered here, such diagrams could only appear from fourth order in $\alpha$ on, which is why we do not work with them in this article. In addition, and as we would expect, it is already known that there is an exponentially suppressed difference between scalar field masses in infinite and finite volumes; see, for example, Ref.~\cite{Luscher:1985dn}.


\subsection{Probabilities}

Having derived the density matrix elements, we can now compute the probability of a single $\phi$-particle decaying into two $\varphi$-particles. For the infinite volume case, we can use \cite{Burrage:2025xac}
\begin{eqnarray}
    P^\infty_{0;2}(t) &=&  \frac{1}{2} \int d\Pi^\varphi_{\mathbf{p}} d\Pi^\varphi_{\mathbf{k}} 
\rho_{0,2;0,2}(;\mathbf{p}, \mathbf{k}|;\mathbf{p}, \mathbf{k}|t)
\nonumber
\\
&\approx&  
\frac{\alpha^2 \mathcal{M}^2}{8} \int d\Pi^\varphi_{\mathbf{p}} d\Pi^\varphi_{\mathbf{k}} 
\frac{\rho^\infty_{1,0;1,0}(\mathbf{p}+\mathbf{k};|\mathbf{p}+\mathbf{k};|0) t^2}{(E^\phi_{\mathbf{p}+\mathbf{k}})^2}
\mathrm{sinc}^2\left[ \frac{1}{2} (E^\varphi_{\mathbf{p}} + E^\varphi_{\mathbf{k}} - E^\phi_{\mathbf{p}+\mathbf{k}}) t \right]~,
\end{eqnarray}
where $\mathrm{sinc}(x) = \sin(x)/x$. As an example, we choose to work in the rest frame of the single $\phi$-particle, such that we can use $\rho^\infty_{1,0;1,0}(\mathbf{p}+\mathbf{k};|\mathbf{p}+\mathbf{k};|0) = (2\pi)^3 2 E^\phi_{\mathbf{p}+\mathbf{k}} \delta^{(3)}(\mathbf{p}+\mathbf{k})$ for the initial density matrix element. Consequently, we are left with
\begin{eqnarray}
\label{eq:ProbCont12}
    P^\infty_{0;2}(t) 
&\approx&  
 \frac{\alpha^2 \mathcal{M}^2}{16M} \int \frac{d^3p}{(2\pi)^3  }   
\frac{t^2}{ (E^\varphi_{\mathbf{p}})^2}
\mathrm{sinc}^2\left[ \frac{1}{2} (2E^\varphi_{\mathbf{p}}  - M) t \right]
~.
\end{eqnarray}
Similarly, we find for the resting $\phi$-particle confined to a volume $V = L^3$:
\begin{eqnarray}
\label{eq:ProbDisc12}
    P^V_{0;2}(t) 
    &\approx&  
    \frac{\alpha^2 \mathcal{M}^2}{32} 
    \frac{1}{V^2}\sum_{\mathbf{p}\mathbf{k}}\frac{1}{   E^\varphi_\mathbf{p}E^\varphi_\mathbf{k}}
\frac{\rho^V_{1,0;1,0}(\mathbf{p}+\mathbf{k};|\mathbf{p}+\mathbf{k};|0) t^2}{(E^\phi_{\mathbf{p}+\mathbf{k}})^2}
\mathrm{sinc}^2\left[ \frac{1}{2} (E^\varphi_{\mathbf{p}} + E^\varphi_{\mathbf{k}} - E^\phi_{\mathbf{p}+\mathbf{k}}) t \right]
\nonumber
\\
&\approx&  
    \frac{\alpha^2 \mathcal{M}^2 }{16M} 
    \frac{1}{V}\sum_{\mathbf{p}}
\frac{t^2}{ (E^\varphi_{\mathbf{p}})^2}
\mathrm{sinc}^2\left[ \frac{1}{2} (2E^\varphi_{\mathbf{p}}  - M) t \right]
\nonumber
\\
    &\approx&  
\frac{\alpha^2 \mathcal{M}^2 }{16M} 
    \frac{1}{V}\sum_{n \in \mathbb{Z}^+}
\frac{ \aleph_n t^2}{ (E^\varphi_{\mathbf{p}})^2}
\mathrm{sinc}^2\left[ \frac{1}{2} (2E^\varphi_{\mathbf{p}}  - M) t \right]  
~,
\end{eqnarray}
where we have used ${\rho^V_{1,0;1,0}(\mathbf{p}+\mathbf{k};|\mathbf{p}+\mathbf{k};|0) =  2 E^\phi_{\mathbf{p}+\mathbf{k}}  V \delta_{\mathbf{p}+\mathbf{k},\mathbf{0}}}$, and $|\mathbf{p}| =: 2\pi \sqrt{n}/L$ with $\aleph_n$ being the number of integer vectors $\mathbf{z}$ that fulfill $\mathbf{z}^2 = n$ \cite{Lellouch:2000pv}.

Actually computing the remaining integral in Eq.~(\ref{eq:ProbCont12}) and the sum in Eq.~(\ref{eq:ProbDisc12}) is rather challenging. Therefore, we instead extract the respective terms corresponding to energy conservation, i.e., for which $|\mathbf{p}| = \sqrt{M^2/4 - m_\varphi^2}$ and $\mathrm{sinc}\left[ \frac{1}{2} (2E^\varphi_{\mathbf{p}}  - M) t \right] \to 1$. While there are no complications for Eq.~(\ref{eq:ProbDisc12}) when extracting this term since each summand is already dimensionless, we must introduce an unknown factor $\mathcal{C}$ with $[\mathcal{C}] =3$ as a replacement for the differentials in Eq.~(\ref{eq:ProbCont12}). Consequently, we obtain
\begin{eqnarray}
    P^\infty_{0;2}\left(|\mathbf{p}| ;t\right) 
&\approx&  
 \frac{\alpha^2 \mathcal{M}^2}{16M} \frac{ \mathcal{C}}{(2\pi)^3  }   
\frac{t^2}{ (E^\varphi_{\mathbf{p}})^2}
~,
\\
    P^V_{0;2}\left(|\mathbf{p}| ;t\right) 
&\approx&  
\frac{\alpha^2 \mathcal{M}^2 }{16M} 
    \frac{1}{V}
\frac{ \aleph_n t^2}{ (E^\varphi_{\mathbf{p}})^2}~.
\end{eqnarray}
Note that these two expressions are only valid for times that comply with the permitted range of probabilities. If we compare both results, we find
\begin{eqnarray}
\label{eq:LLfactor}
    \frac{P^\infty_{0;2}\left(|\mathbf{p}| ;t\right) }{P^V_{0;2}\left(|\mathbf{p}| ;t\right)}
    &\approx&
    \frac{\mathcal{C} V}{(2\pi)^3\aleph_n }~.
\end{eqnarray}
Since it essentially stems from the same type of decay process, the ratio in Eq.~(\ref{eq:LLfactor}) must actually be the Lellouch-L\"uscher factor that was derived in Ref.~\cite{Lellouch:2000pv} for a resting scalar decaying into two copies of another scalar, while neglecting self-interactions of the second scalar, and which relates the decay amplitudes in the infinite and finite volume cases \cite{Pang:2023jri}. From this, we conclude that $\mathcal{C} = 4(2\pi)^3M^3$, such that we finally arrive at 
\begin{eqnarray}
\label{eq:2InfFinComp}
    \frac{P^\infty_{0;2}\left(|\mathbf{p}| ;t\right) }{P^V_{0;2}\left(|\mathbf{p}| ;t\right)}
    &\approx&
    \frac{4 M^3 V}{\aleph_n}~.
\end{eqnarray}
Now we can clearly see that, depending on the values of $M$, $V$ and $n$ (the last of which is determined through $m_\varphi$), there can be massive differences between probabilities in the infinite and finite volume cases. To fully assess these differences, however, we would be required to evaluate the full integral and sum in Eqs.~(\ref{eq:ProbCont12}) and (\ref{eq:ProbDisc12}), which is beyond the scope of the current article. 


\section{Neutron decay toy model}
\label{sec:neutrons}

Next, we will discuss a simple toy model of neutron decay. More precisely, we consider a scalar field $\phi$ with mass $M$ as a 'neutron' that decays into three other scalar fields $\varphi$ (a 'proton'), $\chi$ (an 'electron') and $\nu$ (an 'anti-electron neutrino') with masses $m_\varphi$, $m_\chi$ and $m_\nu$, respectively, via a contact interaction
\begin{eqnarray}
    S_{\text{int}}[\phi;\varphi;\chi;\nu] 
    &=&
    \int_{x\in\Omega_{t,V}} 
    \left[ -\alpha \phi \varphi \chi \nu   \right]~,
\end{eqnarray}
where, again, $\alpha \ll 1$ is a dimensionless coupling constant. Apart from this interaction, all fields are free with $\phi$ and $\varphi$ having the same free actions as in Eq.~(\ref{eq:phivarphiac}), and the other scalars have
\begin{eqnarray}
    S_\chi[\chi] &=& \int_x \left[ -\frac{1}{2}(\partial \chi)^2 -\frac{1}{2} m_\chi^2\chi^2 \right]~,~~~
      S_\nu[\nu] = \int_x \left[ -\frac{1}{2}(\partial \nu)^2 -\frac{1}{2} m_\nu^2\nu^2 \right]~,
\end{eqnarray}
such that the total action is given by $S[\phi;\varphi;\chi;\nu] =S_\phi[\phi]  + S_\varphi[\varphi] +S_\chi[\chi]  + S_\nu[\nu]+ S_{\text{int}}[\phi;\varphi;\chi;\nu] $. We take the neutron and the produced proton to both be at rest, and, in order to properly describe the contact interaction, we choose $\alpha = G_F (M-m_\varphi)^2v_{ud}$, where $G_F = 1.16637 \times 10^{-5}$ $\text{GeV}^{-2}$ is the Fermi coupling constant and $v_{ud} = 0.97367$ \cite{PDG:2024cfk} is the first entry of the Cabibbo-Kobayashi-Maskawa (CKM) matrix \cite{Cabibbo:1963yz,Kobayashi:1973fv}. Since $G_F \sim m_W^{-2}$, this choice of coupling constant takes into account the ratio of the energy transferred by the off-shell W boson to its on-shell rest mass $m_W$ that appears in the W boson propagator. For the masses, we use the values $M = 939.5654205$ MeV, $m_\varphi = 938.2720881$ MeV, $m_\chi = 0.5109980$ MeV \cite{PDG:2024cfk}, and $m_\nu \approx 0.7$ eV.

Since neutrons, protons and electrons are usually at least partially confined in neutron lifetime experiments, we propose that such confinement can render the measured neutron lifetimes to be sensitive to confining volumes and, generally, dependent on the boundary conditions given by the experimental environment the neutron decay is studied in. Using the neutron decay toy model introduced above, we will investigate our proposal of a volume dependence. For this, we will consider a neutron to be confined in a magnetic trap of the same form as in the upcoming $\tau$SPECT experiment \cite{Auler:2023tuf,Auler:2025mbm}, i.e., a cylinder with an approximate volume of $V = (50 \,\text{mm}/2)^2 \cdot \pi \cdot 1 \,\text{m} $. What makes this setup particularly appealing for us are the circumstances that neutrons (and most of their decay products) are confined in all three spatial directions and get perfectly reflected on the boundaries, which allows us to work with a discrete momentum basis and use the formulas introduced in Sec.~\ref{sec:pre}. A deviation from almost perfect reflectivity or a change in the confinement structure of the trap, for example, if it was half open in one direction or if there was a certain amount of leakage of the neutron and its daughter particles, would lead to a drastic modification of our computation. Consequently, we suggest that such differences can potentially explain the variance of neutron lifetimes determined by experiments and contribute to the resolution of the neutron lifetime problem.

For our investigation, we will, at first, follow the same procedure as in Sec.~\ref{sec:two}, i.e, we will compute the density matrices describing the decay for the infinite and finite volume cases and then obtain the probabilities. For the finite volume case, we will make the naive assumption that all considered particles are confined within the trap. From this we will find an unreasonably huge value for $\tau$. Therefore, we will refine the computation by considering that the anti-electron neutrino $\nu$ is essentially unconfined and should be treated as having a continuous spectrum. In this way, the resulting neutron lifetime will be improved by multiple orders of magnitude. Though, this will still not be close enough to experimental results, but point us to the necessity of considering correlations between the neutron and its decay products. Taking this into account, we will be able to get impressively close to experimentally determined neutron lifetimes. Since we will have used the volume of the magnetic trap in the $\tau$SPECT experiment as a parameter in our computation, we interpret this as supporting evidence for our proposal of a volume dependence for the neutron lifetime. 


\subsection{Density matrix elements}

\begin{figure}[htbp]
\begin{center}
\includegraphics[scale=0.6]{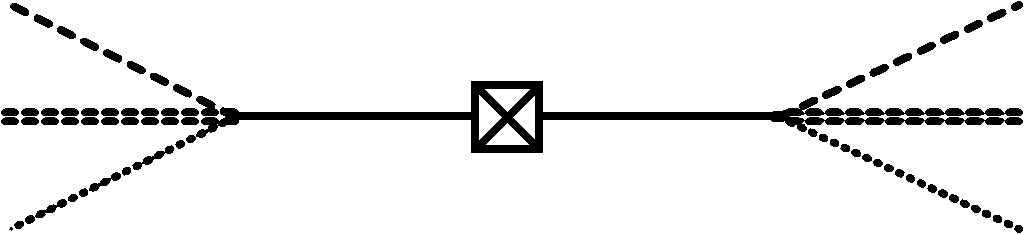}
\caption{Diagram for neutron decay; the crossed box represents the single $\phi$-particle density matrix elements at the initial time $0$, a solid line is a $\phi$-propagator, a double dotted line stands for a $\varphi$-propagator, a dotted line represents a $\chi$-propagator, and a small dotted line depicts a $\nu$-propagator. To the left and to the right of the crossed box, time evolves from $0$ to the final time $t$. More precisely, the left-hand side of this diagram shows the evolution $\bra{\mathbf{p}+\mathbf{k}+\mathbf{l};;;;0} \to  \bra{;\mathbf{p}; \mathbf{k};\mathbf{l};t}$ and the right-hand side depicts $\ket{\mathbf{p}'+\mathbf{k}'+\mathbf{l}';;;;0} \to \ket{;\mathbf{p}'; \mathbf{k}';\mathbf{l}';t}$.}
\label{fig:decay}
\end{center}
\end{figure}

We want to compute the density matrix elements $\rho^\infty_{0,1,1,1;0,1,1,1}(;\mathbf{p};\mathbf{k};\mathbf{l}|;\mathbf{p}';\mathbf{k}';\mathbf{l}'|t)$ up to second order in $\alpha$ under the assumption that only $\rho^\infty_{1,0,0,0;1,0,0,0}(\mathbf{q};;;|\mathbf{q}';;;|0)$ is non-vanishing at the initial time, i.e., there was only a single neutron. We ignore all disconnected diagrams and only consider the one shown in Fig.~\ref{fig:decay}. Using the method presented in Ref.~\cite{Kading:2022hhc}, we have to compute:
\begin{eqnarray}\label{eq:12Scatt}
&&\rho^\infty_{0,1,1,1;0,1,1,1}(;\mathbf{p};\mathbf{k};\mathbf{l}|;\mathbf{p}';\mathbf{k}';\mathbf{l}'|t) 
\nonumber
\\
&&
~~~~~
\approx
\alpha^2
\lim_{\substack{x^{0(')}_{\varphi,\chi,\nu}\,\to\, t^{+}\\y^{0(')}\,\to\, 0^-}}
\int d\Pi^\phi_{\mathbf{q}}d\Pi^\phi_{\mathbf{q}'}\rho^\infty_{1,0,0,0;1,0,0,0}(\mathbf{q};;;|\mathbf{q}';;;|0)
\nonumber
\\
&&
~~~~~~~~
\times
\int_{\mathbf{x}_{\varphi}\mathbf{x}'_{\varphi}\mathbf{x}_{\chi}\mathbf{x}'_{\chi}\mathbf{x}_{\nu}\mathbf{x}'_{\nu}\mathbf{y}\mathbf{y}'} 
e^{-\mathrm{i}(\mathbf{p}\cdot\mathbf{x}_{\varphi} + \mathbf{k}\cdot\mathbf{x}_{\chi} + \mathbf{l}\cdot\mathbf{x}_{\nu} - \mathbf{p}'\cdot\mathbf{x}'_{\varphi} - \mathbf{k}'\cdot\mathbf{x}'_{\chi} - \mathbf{l}'\cdot\mathbf{x}'_{\nu})+\mathrm{i}(\mathbf{q}\cdot\mathbf{y}-\mathbf{q}'\cdot\mathbf{y}')}
\nonumber
\\
&&~~~~~~~~
\times
\partial_{x_{\varphi}^0,E^\varphi_{\mathbf{p}}}
\partial_{x_{\varphi}^{0'},E^\varphi_{\mathbf{p}'}}^*
\partial_{x^0_{\chi},E^\chi_{\mathbf{k}}}
\partial_{x^{0'}_{\chi},E^\chi_{\mathbf{k}'}}^*
\partial_{x^0_{\nu},E^\nu_{\mathbf{l}}}
\partial_{x^{0'}_{\nu},E^\nu_{\mathbf{l}'}}^*
\partial_{y^0,E^\phi_{\mathbf{q}}}^*
\partial_{y^{0'},E^\phi_{\mathbf{q}'}}
\nonumber
\\
&&~~~~~~~~
\times
\int\mathcal{D}\phi^{\pm}\mathcal{D}\varphi^{\pm}\mathcal{D}\chi^{\pm} \mathcal{D}\nu^{\pm} e^{\mathrm{i}\widehat{S}_{\phi}[\phi]+\mathrm{i}\widehat{S}_{\varphi}[\varphi]+\mathrm{i}\widehat{S}_{\chi}[\chi]+\mathrm{i}\widehat{S}_{\nu}[\nu]}
\nonumber
\\
&&~~~~~~~~
\times
\varphi^+_{x_{\varphi}}\varphi^-_{x'_{\varphi}}
\chi^+_{x_{\chi}}\chi^-_{x'_{\chi}}
\nu^+_{x_{\nu}}\nu^-_{x'_{\nu}}
\int_{zz'}
\phi^+_{z}\phi^-_{z'}
\varphi^+_{z}\varphi^-_{z'}
\chi^+_{z}\chi^-_{z'}
\nu^+_{z}\nu^-_{z'}
\phi^+_{y}\phi^-_{y'}~,
\end{eqnarray}
where $\partial_{x_{\varphi}^0,E^\varphi_{\mathbf{p}}} := \partial_{x_{\varphi}^0} - \mathrm{i}E^\varphi_{\mathbf{p}}$, $\mathcal{D}\phi^{\pm} := \mathcal{D}\phi^{+}\mathcal{D}\phi^{-}$, and $\widehat{S}_\phi[\phi] := S_\phi[\phi^+] - S_\phi[\phi^-]$; see App.~\ref{app:SK} for an example that explains how to arrive at such an equation. Note that the path integrals in Eq.~(\ref{eq:12Scatt}) can only lead to contractions of two $+$ or two $-$ labeled fields \cite{Kading:2022jjl}, such that we have Feynman (F) or Dyson (D) propagators $D^{\mathrm{F},\mathrm{D}}$, $\Delta^{\varphi,\mathrm{F},\mathrm{D}}$, $\Delta^{\chi,\mathrm{F},\mathrm{D}}$ and $\Delta^{\nu,\mathrm{F},\mathrm{D}}$ for $\phi$, $\varphi$, $\chi$ and $\nu$, respectively. After evaluating the path integrals, we are left with
\begin{eqnarray}
&&\rho^\infty_{0,1,1,1;0,1,1,1}(;\mathbf{p};\mathbf{k};\mathbf{l}|;\mathbf{p}';\mathbf{k}';\mathbf{l}'|t) 
\nonumber
\\
&&~~~~~
\approx
\alpha^2
\lim_{\substack{x^{0(')}_{\varphi,\chi,\nu}\,\to\, t^{+}\\y^{0(')}\,\to\, 0^-}}
\int d\Pi_{\mathbf{q}}^\phi d\Pi_{\mathbf{q}'}^\phi \rho^\infty_{1,0,0,0;1,0,0,0}(\mathbf{q};;;|\mathbf{q}';;;|0)
\nonumber
\\
&&~~~~~~~~
\times
\int_{\mathbf{x}_{\varphi}\mathbf{x}'_{\varphi}\mathbf{x}_{\chi}\mathbf{x}'_{\chi}\mathbf{x}_{\nu}\mathbf{x}'_{\nu}\mathbf{y}\mathbf{y}'} 
e^{-\mathrm{i}(\mathbf{p}\cdot\mathbf{x}_{\varphi} + \mathbf{k}\cdot\mathbf{x}_{\chi} + \mathbf{l}\cdot\mathbf{x}_{\nu} - \mathbf{p}'\cdot\mathbf{x}'_{\varphi} - \mathbf{k}'\cdot\mathbf{x}'_{\chi} - \mathbf{l}'\cdot\mathbf{x}'_{\nu})+\mathrm{i}(\mathbf{q}\cdot\mathbf{y}-\mathbf{q}'\cdot\mathbf{y}')}
\nonumber
\\
&&~~~~~~~~
\times
\partial_{x_{\varphi}^0,E^\varphi_{\mathbf{p}}}
\partial_{x_{\varphi}^{0'},E^\varphi_{\mathbf{p}'}}^*
\partial_{x^0_{\chi},E^\chi_{\mathbf{k}}}
\partial_{x^{0'}_{\chi},E^\chi_{\mathbf{k}'}}^*
\partial_{x^0_{\nu},E^\nu_{\mathbf{l}}}
\partial_{x^{0'}_{\nu},E^\nu_{\mathbf{l}'}}^*
\partial_{y^0,E^\phi_{\mathbf{q}}}^*
\partial_{y^{0'},E^\phi_{\mathbf{q}'}}
\nonumber
\\
&&~~~~~~~~
\times
\int_{zz'}
\Delta^{\varphi,\mathrm{F}}_{x_{\varphi}z}
\Delta^{\varphi,\mathrm{D}}_{x'_{\varphi}z'}
\Delta^{\chi,\mathrm{F}}_{x_{\chi}z}
\Delta^{\chi,\mathrm{D}}_{x'_{\chi}z'}
\Delta^{\nu,\mathrm{F}}_{x_{\nu}z}
\Delta^{\nu,\mathrm{D}}_{x'_{\nu}z'}
D^\mathrm{F}_{zy} D^\mathrm{D}_{z'y'}~.
\end{eqnarray}
Finally, after substituting explicit expressions for the propagators, we can evaluate the momentum integrals by using Eq.~(\ref{eq:AppXint}) and Cauchy's integral formula, such that we arrive at
\begin{eqnarray}
\label{eq:NeutFirstresutlinfi}
&&\rho^\infty_{0,1,1,1;0,1,1,1}(;\mathbf{p};\mathbf{k};\mathbf{l}|;\mathbf{p}';\mathbf{k}';\mathbf{l}'|t) 
\nonumber
\\
&&~~~~~
\approx
\frac{\alpha^2 \rho^\infty_{1,0,0,0;1,0,0,0}(\mathbf{p} + \mathbf{k} + \mathbf{l};;;|\mathbf{p}' + \mathbf{k}' +\mathbf{l}';;;|0) }{4 E^\phi_{\mathbf{p}+\mathbf{k}+\mathbf{l}}E^\phi_{\mathbf{p}'+\mathbf{k}'+\mathbf{l}'}(E^\phi_{\mathbf{p}+\mathbf{k}+\mathbf{l}}-E^\varphi_\mathbf{p} - E^\chi_\mathbf{k} - E^\nu_\mathbf{l})(E^\phi_{\mathbf{p}'+\mathbf{k}'+\mathbf{l}'}-E^\varphi_{\mathbf{p}'} - E^\chi_{\mathbf{k}'} - E^\nu_{\mathbf{l}'})}
 \nonumber
\\
&&~~~~~~~~
\times
\bigg[  
e^{-\mathrm{i}(E^\varphi_{\mathbf{p}} + E^\chi_{\mathbf{k}}+ E^\nu_{\mathbf{l}})t}
-
e^{-\mathrm{i}E^\phi_{\mathbf{p}+\mathbf{k}+\mathbf{l}}t}
\bigg]
\bigg[  
e^{\mathrm{i}(E^\varphi_{\mathbf{p}'} + E^\chi_{\mathbf{k}'}+ E^\nu_{\mathbf{l}'})t}
-
e^{\mathrm{i}E^\phi_{\mathbf{p}'+\mathbf{k}'+\mathbf{l}'}t}
\bigg]
~.
\end{eqnarray}
Using the same formalism as in Sec.~\ref{sec:DME12}, we find that also for this example, the density matrices for the infinite and finite volume cases are essentially the same (when ignoring mass corrections), apart from the differing initial density matrix elements, i.e., 
\begin{eqnarray}
&&\rho^V_{0,1,1,1;0,1,1,1}(;\mathbf{p};\mathbf{k};\mathbf{l}|;\mathbf{p}';\mathbf{k}';\mathbf{l}'|t) 
\nonumber
\\
&&~~~~~
\approx
\frac{\alpha^2 \rho^V_{1,0,0,0;1,0,0,0}(\mathbf{p} + \mathbf{k} + \mathbf{l};;;|\mathbf{p}' + \mathbf{k}' +\mathbf{l}';;;|0) }{4 E^\phi_{\mathbf{p}+\mathbf{k}+\mathbf{l}}E^\phi_{\mathbf{p}'+\mathbf{k}'+\mathbf{l}'}(E^\phi_{\mathbf{p}+\mathbf{k}+\mathbf{l}}-E^\varphi_\mathbf{p} - E^\chi_\mathbf{k} - E^\nu_\mathbf{l})(E^\phi_{\mathbf{p}'+\mathbf{k}'+\mathbf{l}'}-E^\varphi_{\mathbf{p}'} - E^\chi_{\mathbf{k}'} - E^\nu_{\mathbf{l}'})}
 \nonumber
\\
&&~~~~~~~~
\times
\bigg[  
e^{-\mathrm{i}(E^\varphi_{\mathbf{p}} + E^\chi_{\mathbf{k}}+ E^\nu_{\mathbf{l}})t}
-
e^{-\mathrm{i}E^\phi_{\mathbf{p}+\mathbf{k}+\mathbf{l}}t}
\bigg]
\bigg[  
e^{\mathrm{i}(E^\varphi_{\mathbf{p}'} + E^\chi_{\mathbf{k}'}+ E^\nu_{\mathbf{l}'})t}
-
e^{\mathrm{i}E^\phi_{\mathbf{p}'+\mathbf{k}'+\mathbf{l}'}t}
\bigg]
~.
\end{eqnarray}
The reasons for the similarity of both results are the same as those that we have explained below Eq.~(\ref{eq:disc12deres}). In particular, also for the considered neutron decay toy model, mass correcting loop diagrams, which would lead to different results in infinite and finite volumes, can at earliest appear at $\mathcal{O}(\alpha^4)$.

\subsection{Probabilities}
\label{sec:NeutrProb}

We can now compute the probabilities for finding a proton, an electron, and an anti-electron neutrino at time $t$ after having had a single neutron at the initial time $0$. For the infinite volume case, we find
\begin{eqnarray}
    P^\infty_{0;1;1;1}(t) &=&  \int d\Pi^\varphi_{\mathbf{p}} d\Pi^\chi_{\mathbf{k}} d\Pi^\nu_{\mathbf{l}}
\rho^\infty_{0,1,1,1;0,1,1,1}(;\mathbf{p};\mathbf{k};\mathbf{l}|;\mathbf{p};\mathbf{k};\mathbf{l}|t) 
\nonumber
\\
&\approx&
\frac{\alpha^2}{4}\int d\Pi^\varphi_{\mathbf{p}} d\Pi^\chi_{\mathbf{k}} d\Pi^\nu_{\mathbf{l}}
\frac{ \rho_{1,0,0,0;1,0,0,0}(\mathbf{p} + \mathbf{k} + \mathbf{l};;;|\mathbf{p} + \mathbf{k} +\mathbf{l};;;|0)t^2 }{ (E^\phi_{\mathbf{p}+\mathbf{k}+\mathbf{l}})^2 }
 \nonumber
\\
&&
~~~~~~~
\times
\mathrm{sinc}^2\left[ \frac{1}{2} (E^\varphi_\mathbf{p} + E^\chi_\mathbf{k} + E^\nu_\mathbf{l} - E^\phi_{\mathbf{p}+\mathbf{k}+\mathbf{l}}) t \right]
~.
\end{eqnarray}
We choose $\rho^\infty_{1,0,0,0;1,0,0,0}(\mathbf{p} + \mathbf{k} + \mathbf{l};;;|\mathbf{p} + \mathbf{k} +\mathbf{l};;;|0) = (2\pi)^3 2 E^\phi_{\mathbf{p} + \mathbf{k} +\mathbf{l}} \delta^{(3)}(\mathbf{p} + \mathbf{k} +\mathbf{l})$ for a neutron in its rest frame, and we obtain
\begin{eqnarray}
    P^\infty_{0;1;1;1}(t) 
&\approx&
\frac{\alpha^2}{4}\int  d\Pi^\varphi_{\mathbf{p}} d\Pi^\nu_{\mathbf{l}}
\frac{ t^2 }{ M E^\chi_{\mathbf{p}+\mathbf{l}} }
\mathrm{sinc}^2\left[ \frac{1}{2} (E^\varphi_{\mathbf{p}} + E^\chi_{\mathbf{p}+\mathbf{l}} + E^\nu_\mathbf{l} - M) t \right]
~.
\end{eqnarray}
As a further simplification, we consider the case in which the proton is also almost at rest, such that $E^\varphi_p \approx m_\varphi$. Consequently, we must approximate the integral over the proton momentum. For this, we do the same replacement as in Sec.~\ref{sec:two}, i.e., $\int d^3p \to \mathcal{C} = 4(2\pi)^3 M^3$. Certainly, we can only speculate that this replacement is applicable here as well since the neutron toy model decay process is different from the one discussed in Sec.~\ref{sec:two}. Therefore, we incorporate this assumption into our toy model and will see whether we will obtain a sensible result. After this approximation, we are left with
\begin{eqnarray}
    P^\infty_{0;1;1;1}(t) 
&\approx&
\frac{\alpha^2}{2}\int d\Pi^\nu_{\mathbf{l}}
\frac{ M^2 t^2 }{  m_\varphi E^\chi_\mathbf{l} }
\mathrm{sinc}^2\left[ \frac{1}{2} (m_\varphi + E^\chi_\mathbf{l} + E^\nu_\mathbf{l} - M) t \right]
~.
\end{eqnarray}
Finally, we restrict us to the case of energy conservation and replace $\int d^3l \to \mathcal{C} = 4(2\pi)^3 M^3$ for 
\begin{eqnarray}
\label{eq:Moml}
    |\mathbf{l}| = \frac{1}{2(M-m_\varphi)}
    \sqrt{
    \prod\limits_{a,b=\pm} (M-m_\varphi+am_\chi+ b m_\nu)
    }~,
\end{eqnarray}
such that
\begin{eqnarray}
    P^\infty_{0;1;1;1}(|\mathbf{l}|;t) 
&\approx&
\frac{\alpha^2 M^5 t^2 }{  m_\varphi E^\chi_\mathbf{l} E^\nu_\mathbf{l} }
~.
\end{eqnarray}
Next, we consider the case of a finite volume:
\begin{eqnarray}
    P^V_{0;1;1;1}(t) 
&\approx&
\frac{\alpha^2}{32V^3}\sum_{\mathbf{p}\mathbf{k}\mathbf{l}}\frac{1}{   E^\varphi_\mathbf{p} E^\chi_\mathbf{k} E^\nu_\mathbf{l}}
\frac{ \rho^V_{1,0,0,0;1,0,0,0}(\mathbf{p} + \mathbf{k} + \mathbf{l};;;|\mathbf{p} + \mathbf{k} +\mathbf{l};;;|0) t^2 }{ (E^\phi_{\mathbf{p}+\mathbf{k}+\mathbf{l}})^2 }
 \nonumber
\\
&&
~~~~~~~~~~~~
\times
\mathrm{sinc}^2\left[ \frac{1}{2} (E^\varphi_\mathbf{p} + E^\chi_\mathbf{k} + E^\nu_\mathbf{l} - E^\phi_{\mathbf{p}+\mathbf{k}+\mathbf{l}}) t \right]
~.
\end{eqnarray}
Choosing $\rho^V_{1,0,0,0;1,0,0,0,}(\mathbf{p} + \mathbf{k} + \mathbf{l};;;|\mathbf{p} + \mathbf{k} + \mathbf{l};;;|0) = 2 E^\phi_{\mathbf{p} + \mathbf{k} + \mathbf{l}}  V \delta_{\mathbf{p} + \mathbf{k} + \mathbf{l},\mathbf{0}}$
and considering a resting proton, we arrive at
\begin{eqnarray}
\label{eq:NeutronVrest}
    P^V_{0;1;1;1}(t) 
&\approx&
\frac{\alpha^2}{16V^2}\sum_{\mathbf{l}}
\frac{  t^2 }{   M m_\varphi E^\chi_{\mathbf{l}} E^\nu_\mathbf{l}}
\mathrm{sinc}^2\left[ \frac{1}{2} (m_\varphi + E^\chi_{\mathbf{l}} + E^\nu_\mathbf{l} - M) t \right]
\nonumber
\\
&\approx&
\frac{\alpha^2}{16V^2}\sum_{n \in \mathbb{Z}^+}
\frac{ \aleph_n t^2 }{   M m_\varphi E^\chi_{\mathbf{l}} E^\nu_\mathbf{l}}
\mathrm{sinc}^2\left[ \frac{1}{2} (m_\varphi + E^\chi_{\mathbf{l}} + E^\nu_\mathbf{l} - M) t \right]
~.
\end{eqnarray}
After extracting the term corresponding to energy conservation, we find
\begin{eqnarray}
\label{eq:FinProbNaive}
    P^V_{0;1;1;1}(|\mathbf{l}|;t) 
&\approx&
\frac{\alpha^2}{16V^2}
\frac{ \aleph_n t^2 }{   M m_\varphi E^\chi_{\mathbf{l}} E^\nu_\mathbf{l}}
~.
\end{eqnarray}
Note that the $\text{sinc}^2$-function in Eq.~(\ref{eq:NeutronVrest}) has its maximum at the value of $|\mathbf{l}|$ given in Eq.~(\ref{eq:Moml}) and then falls off rapidly, which makes Eq.~(\ref{eq:FinProbNaive}) a sensible approximation. Akin to Eq.~(\ref{eq:2InfFinComp}), the comparison between the probabilities in both cases gives us
\begin{eqnarray}
\label{eq:NeuInffincomp}
\frac{P^\infty_{0;1;1;1}(|\mathbf{l}|;t)}{P^V_{0;1;1;1}(|\mathbf{l}|;t)}
&\approx&
\frac{16 M^6 V^2 }{\aleph_n}
~.
\end{eqnarray}
Next, we will check whether a reasonable neutron lifetime can be derived from Eq.~(\ref{eq:FinProbNaive}). For this, we consider that the survival probability of a neutron at time $t$ is, at the considered order in the coupling constant, essentially equal to $1-  P^V_{0;1;1;1}(|\mathbf{l}|;t) $. Since the mean lifetime of a particle is defined as the time $\tau$ at which the particle's survival probability has dropped to $1/e$, we obtain
\begin{eqnarray}
\label{eq:taunaive}
\tau &=& \frac{4 V}{G_F (M-m_\varphi)^2 v_{ud}} \sqrt{ \frac{ \left(  1 -\frac{1}{e}  \right) M m_\varphi E^\chi_{\mathbf{l}} E^\nu_\mathbf{l} }{\aleph_n} }
\approx
\frac{1.68\cdot 10^{28}}{\sqrt{\aleph_n}}\,\text{s}
~.
\end{eqnarray}
Since $\aleph_n$ is difficult to calculate for large numbers, we make a rough overestimation in order to demonstrate that it can not be sufficiently large to get the result in Eq.~(\ref{eq:taunaive}) close to experimental values of the neutron lifetime. For large $n$, we can comfortably say that $\aleph_n \leq n$. Therefore, for our overestimation, we set $\aleph_n \equiv n$. For simplicity and only for this overestimation, we consider the confining volume to be a cube with $L = 1\,\text{m}$ rather than a cylinder. Within such a cube, we would find 
\begin{eqnarray}
    \sqrt{\aleph_n} &=&\frac{L |\mathbf{l}|}{2\pi} \approx 4.4\times 10^{11} 
\end{eqnarray}
with $|\mathbf{l}|$ taken from Eq.~(\ref{eq:Moml}). Substituting this largely overestimated result into Eq.~(\ref{eq:taunaive}), the value of $\tau$ is still more than twelve orders of magnitude larger than the experimentally found values for the neutron lifetime. Consequently, we will have to improve the toy model in order to get closer to realistic results.


\subsection{More realistic model: unconfined neutrino}
\label{sec:unconeutrino}

A first step for improving our model is taking into account that we should not treat the neutrino as being confined within the finite volume. This means that the neutrino actually has a continuous momentum spectrum. In this case, the probability of finding the decay products of a neutron is
\begin{eqnarray}
    P_{0;1;1;1}(t) 
&\approx&
\frac{\alpha^2}{32V^2}\sum_{\mathbf{p}\mathbf{k}}\int\frac{d^3l}{ (2\pi)^3  E^\varphi_\mathbf{p} E^\chi_\mathbf{k} E^\nu_\mathbf{l}}
\frac{ \rho_{1,0,0,0;1,0,0,0}(\mathbf{p} + \mathbf{k} + \mathbf{l};;;|\mathbf{p} + \mathbf{k} +\mathbf{l};;;|0)t^2 }{ (E^\phi_{\mathbf{p}+\mathbf{k}+\mathbf{l}})^2 }
 \nonumber
\\
&&
~~~~~~~~~~~~~~~
\times
\mathrm{sinc}^2\left[ \frac{1}{2} (E^\varphi_\mathbf{p} + E^\chi_\mathbf{k} + E^\nu_\mathbf{l} - E^\phi_{\mathbf{p}+\mathbf{k}+\mathbf{l}}) t \right]
~.
\end{eqnarray}
We expect this improvement to lead to a neutron lifetime closer to the measured values for two reasons. First of all, with this improvement, the model is much more realistic, which should naturally lead to a better prediction. Secondly, in Eq.~(\ref{eq:NeuInffincomp}), we have already seen that computing the probability at time $t$ in an infinite volume will give a significantly larger result than in a finite volume. It seems natural that this statement also holds if only one of the degrees of freedom, i.e., here the anti-neutrino, is considered in an infinite volume instead of a finite one. For the improved model, this would mean that the predicted decay probability at time $t$ will be larger than for the model in Sec.~\ref{sec:NeutrProb}. Consequently, this would imply a shorter predicted neutron lifetime. Since the neutron itself is still confined, we again use $\rho_{1,0,0,0;1,0,0,0,}({\mathbf{p}+\mathbf{k}+\mathbf{l}};;;|{\mathbf{p}+\mathbf{k}+\mathbf{l}};;;|0) = 2 E^\phi_{\mathbf{p}+\mathbf{k}+\mathbf{l}} V \delta_{{\mathbf{p}+\mathbf{k}+\mathbf{l}},\mathbf{0}}$. Furthermore, we assume a resting proton, such that
\begin{eqnarray}
    P_{0;1;1;1}(t) 
&\approx&
\frac{\alpha^2}{16V}\int\frac{d^3l\,t^2}{ (2\pi)^3  M m_\varphi E^\chi_{ \mathbf{l}} E^\nu_\mathbf{l}}
\mathrm{sinc}^2\left[ \frac{1}{2} (m_\varphi + E^\chi_{\mathbf{l}} + E^\nu_\mathbf{l} - M) t \right]
~.
\end{eqnarray}
For considering the energy conserving case, we replace $\int d^3l \to \mathcal{C} = 4(2\pi)^3 M^3$ and find
\begin{eqnarray}
\label{eq:Neutrinounconfprob}
    P_{0;1;1;1}(|\mathbf{l}|;t) 
&\approx&
\frac{\alpha^2}{4V}\frac{M^2t^2}{    m_\varphi E^\chi_{ \mathbf{l}} E^\nu_\mathbf{l}}
~.
\end{eqnarray}
Therefore, we obtain for the neutron lifetime:
\begin{eqnarray}
\label{eq:taueutrino}
\tau &=& \frac{2}{G_F (M-m_\varphi)^2 v_{ud} M} \sqrt{  \left(  1 -\frac{1}{e}  \right) V m_\varphi E^\chi_{\mathbf{l}} E^\nu_\mathbf{l}  }
\approx
580097.21\,\text{s}
~.
\end{eqnarray}
While this result is still about three orders of magnitude away from the experimentally found values of the neutron lifetime, it is a very strong improvement over what we have found in Eq.~(\ref{eq:taunaive}). In addition, this result points us to another possible upgrade of the toy model since Eq.~(\ref{eq:taueutrino}) can actually be restated as $(2\pi)^3 \sqrt{2\alpha }\tau \approx  886.93~\text{s}$, which is within an interval of a few seconds around the values for the neutron lifetime found in experiments \cite{Wietfeldt:2024oku}. Therefore, it seems that we essentially need to remove one factor of $\alpha$ from the result in Eq.~(\ref{eq:Neutrinounconfprob}) in order to reach this value. In the next subsection, we will discuss how this can be achieved.


\subsection{More realistic model: initial correlations}
\label{sec:initialcorrel}

\begin{figure}[htbp]
\begin{center}
\includegraphics[scale=0.6]{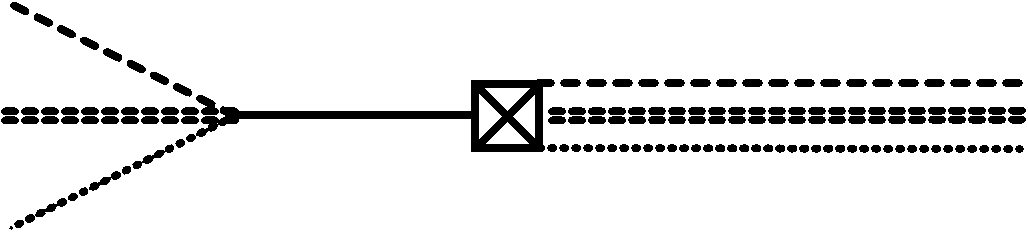}
\caption{Diagram for a neutron correlated with its daughter particles; the crossed box represents the density matrix elements for a correlation between a single $\phi$-particle and its decay products at the initial time $0$. The left-hand side of this diagram shows the evolution $\bra{\mathbf{p}+\mathbf{k}+\mathbf{l};;;;0} \to  \bra{;\mathbf{p}; \mathbf{k};\mathbf{l};t}$ and the right-hand side depicts $\ket{;\mathbf{p}'; \mathbf{k}';\mathbf{l}';0} \to \ket{;\mathbf{p}'; \mathbf{k}';\mathbf{l}';t}$. Note that the conjugated diagram, i.e., the one mirrored along a vertical line through the box, also contributes to Eq.~(\ref{eq:densCorrComp}).}
\label{fig:correlation}
\end{center}
\end{figure}

In the previous subsection, we have seen that a neutron lifetime close to measured values can be predicted if we could work at first order in $\alpha$ instead of at second order. From the diagram in Fig.~\ref{fig:correlation}, we observe that this can be achieved by working with an initial correlation between a neutron and its daughter particles instead of the density matrices for an initially single neutron. Quantum mechanically, an unstable particle decaying into its daughter particles can be interpreted as having a superposition of the particle state $\ket{\text{particle}}$ and a state of its decay products $\ket{\text{products}}$, such that, at the initial time $0$, there is a state $\ket{\Psi(0)} = A(0)\ket{\text{particle}} + B(0)\ket{\text{products}} $ with $|A(0)|^2 \approx 1$ and $|B(0)|^2 \approx 0$, but at a later time $t$, the probability of finding the unstable particle has decreased, i.e., $|A(t)|^2 < |A(0)|^2$, while the probability of finding the daughter particles has increased, i.e., $|B(t)|^2 > |B(0)|^2$. Consequently, it is reasonable for us to assume that there can be a small correlation between a neutron and its decay products even at the initial time, i.e., after the neutron has entered the trap. Computing the decay probability from the initial correlation density matrix elements $\rho_{1,0,0,0;0,1,1,1}(\mathbf{q};;;|;\mathbf{u}';\mathbf{v}';\mathbf{w}'|0)$ is advantageous because they actually describe a process at first order in $\alpha$, which appears to fulfill the requirement we found in Sec.~\ref{sec:unconeutrino} for obtaining a neutron lifetime very close to experimentally derived values. Following the procedure that was exemplified in App.~\ref{app:SK}, for the density matrix elements we find
\begin{eqnarray}
\label{eq:densCorrComp}
\rho_{0,1,1,1;0,1,1,1}(;\mathbf{p};\mathbf{k};\mathbf{l}|;\mathbf{p}';\mathbf{k}';\mathbf{l}'|t) 
&\approx&
\alpha
\frac{\rho_{1,0,0,0;0,1,1,1}(\mathbf{p}+\mathbf{k}+\mathbf{l};;;|;\mathbf{p}';\mathbf{k}';\mathbf{l}'|0)}{2E^\phi_{\mathbf{p}+\mathbf{k}+\mathbf{l}}(E^\varphi_\mathbf{p} +E^\chi_\mathbf{k} +E^\nu_\mathbf{l} -E^\phi_{\mathbf{p}+\mathbf{k}+\mathbf{l}})}
\nonumber
\\
&&
\times
\Bigg[
e^{-\mathrm{i}t(E^\varphi_\mathbf{p} - E^\varphi_{\mathbf{p}'} + E^\chi_\mathbf{k} - E^\chi_{\mathbf{k}'} + E^\nu_\mathbf{l} - E^\nu_{\mathbf{l}'} )}
-
e^{\mathrm{i}t( E^\varphi_{\mathbf{p}'} +  E^\chi_{\mathbf{k}'} + E^\nu_{\mathbf{l}'} - E^\phi_{\mathbf{p}+\mathbf{k}+\mathbf{l}})}
\Bigg]
\nonumber
\\
&&
+[(\mathbf{p},\mathbf{k},\mathbf{l})\longleftrightarrow(\mathbf{p}',\mathbf{k}',\mathbf{l}') ]^\ast~,~~~~
\end{eqnarray}
where we have again only considered the non-divergent diagram in Fig.~(\ref{fig:correlation}) and its conjugated counterpart. Consequently, we obtain the decay probability
\begin{eqnarray}
    P_{0;1;1;1}(t)   
&\approx&
\frac{\alpha}{4V^2}\sum_{\mathbf{p}\mathbf{k}}\int\frac{d^3l}{ (2\pi)^3  E^\varphi_\mathbf{p} E^\chi_\mathbf{k} E^\nu_\mathbf{l}}
\frac{\sin[( E^\varphi_\mathbf{p} +  E^\chi_\mathbf{k} + E^\nu_\mathbf{l} - E^\phi_{\mathbf{p}+\mathbf{k}+\mathbf{l}})t/2]}{E^\phi_{\mathbf{p}+\mathbf{k}+\mathbf{l}}(E^\varphi_\mathbf{p} +E^\chi_\mathbf{k} +E^\nu_\mathbf{l} -E^\phi_{\mathbf{p}+\mathbf{k}+\mathbf{l}})}
\nonumber
\\
&&
\times
\Bigg[
\mathrm{Re}[\rho_{1,0,0,0;0,1,1,1}(\mathbf{p}+\mathbf{k}+\mathbf{l};;;|;\mathbf{p};\mathbf{k};\mathbf{l}|0)] \sin[( E^\varphi_\mathbf{p} +  E^\chi_\mathbf{k} + E^\nu_\mathbf{l} - E^\phi_{\mathbf{p}+\mathbf{k}+\mathbf{l}})t/2]
\nonumber
\\
&&
+
\mathrm{Im}[\rho_{1,0,0,0;0,1,1,1}(\mathbf{p}+\mathbf{k}+\mathbf{l};;;|;\mathbf{p};\mathbf{k};\mathbf{l}|0)] \cos[( E^\varphi_\mathbf{p} +  E^\chi_\mathbf{k} + E^\nu_\mathbf{l} - E^\phi_{\mathbf{p}+\mathbf{k}+\mathbf{l}})t/2]
\Bigg]
~.~~~~~
\end{eqnarray}
If we computed $\rho_{1,0,0,0;0,1,1,1}(\mathbf{p}+\mathbf{k}+\mathbf{l};;;|;\mathbf{p};\mathbf{k};\mathbf{l}|0)$ from a ${\rho_{1,0,0,0;1,0,0,0}(\mathbf{p}+\mathbf{k}+\mathbf{l};;;|\mathbf{p}+\mathbf{k}+\mathbf{l};;;|t')}$ with $t' <0$, then we would find it to be real and proportional to $1/(E^\varphi_\mathbf{p} +  E^\chi_\mathbf{k} + E^\nu_\mathbf{l} - E^\phi_{\mathbf{p}+\mathbf{k}+\mathbf{l}})$. Hence, we assume initial correlation density matrix elements of the form $ {\rho_{1,0,0,0;0,1,1,1}(\mathbf{p}+\mathbf{k}+\mathbf{l};;;|;\mathbf{p};\mathbf{k};\mathbf{l}|0)} =  \mathcal{N}  V \delta_{\mathbf{p} +\mathbf{k} + \mathbf{l},\mathbf{0}}/(E^\varphi_\mathbf{p} +  E^\chi_\mathbf{k} + E^\nu_\mathbf{l} - E^\phi_{\mathbf{p}+\mathbf{k}+\mathbf{l}})$ with some real number $\mathcal{N}$. After also considering a resting proton, we are left with
\begin{eqnarray}
    P_{0;1;1;1}(t)   
&\approx&
\frac{\alpha}{16V}\int\frac{d^3l\,\mathcal{N} t^2}{ (2\pi)^3  M m_\varphi E^\chi_\mathbf{l} E^\nu_\mathbf{l}}
\mathrm{sinc}^2[( m_\varphi +  E^\chi_{\mathbf{l}} + E^\nu_\mathbf{l} - M)t/2]~.
\end{eqnarray}
Finally, after extracting only the energy-conserving term, we arrive at
\begin{eqnarray}
    P_{0;1;1;1}(|\mathbf{l}|;t)   
&\approx&
\frac{\alpha}{4V}\frac{\mathcal{N} M^2 t^2}{ m_\varphi E^\chi_\mathbf{l} E^\nu_\mathbf{l}}~,
\end{eqnarray}
and find 
\begin{eqnarray}
\label{eq:FinalResult}
\tau &=& \frac{2}{\sqrt{\mathcal{N} G_F (M-m_\varphi)^2 v_{ud}} M} \sqrt{  \left(  1 -\frac{1}{e}  \right) V m_\varphi E^\chi_{\mathbf{l}} E^\nu_\mathbf{l}  }
\approx
\frac{2.53}{\sqrt{\mathcal{N}}}\,\text{s}
~.
\end{eqnarray}
 If we choose $\mathcal{N} = 1/ 2(2\pi)^6$, then we obtain $\tau \approx 887.51$ s as was suggested in Sec.~\ref{sec:unconeutrino}.


\section{Conclusions and outlook}
\label{sec:Conclusion}
The neutron lifetime problem remains a great open problem in physics and its resolution would not only impact fundamental parameters of the Standard model but also have implications for cosmology. In this article, we have proposed that the measured neutron lifetime can be dependent on the spatial dimensions of experimental setups, in particular, on volumes within which a neutron is confined. Without the need to introduce any kind of exotic new physics, this proposal might contribute to an explanation of the neutron lifetime discrepancy since individual experiments usually differ in their neutron confinement structures. In order to discuss this proposal, we have considered a rudimentary toy model of decaying neutrons, in which the neutron and its daughter particles are all represented by real scalar particles. As a preparation for computations within this model in infinite and finite volumes, we have introduced some mathematical prerequisites in Sec.~\ref{sec:pre} and, subsequently, in Sec.~\ref{sec:two}, studied differences between infinite and finite volume probabilities for a model of a scalar field $\phi$ decaying into two copies of another scalar field $\varphi$. For this, we have employed the method developed in Ref.~\cite{Kading:2022hhc}, based on the Schwinger-Keldysh formalism \cite{Schwinger,Keldysh}, and tools from thermo field dynamics \cite{Takahasi:1974zn,Arimitsu:1985ez,Arimitsu:1985xm, Khanna}. Using what we have learned in Sec.~\ref{sec:two}, we have computed decay probabilities for neutrons in an infinite volume and in a finite volume akin to the magnetic trap in the upcoming $\tau$SPECT experiment \cite{Auler:2023tuf,Auler:2025mbm}. Initially, we have treated a neutron and all its daughter particles as being confined. However, from this naive calculation we have derived a mean neutron lifetime that was multiple orders of magnitude larger than those measured in experiments. Consequently, we have improved the toy model by taking into account that an anti-electron neutrino would realistically not be confined in such a trap, which means that its momentum spectrum has to be continuous. This change in the model led to a great improvement of the predicted neutron lifetime. Though, our result was still about three orders of magnitude away from actual experimental results. Finally, after also using that the neutron and its daughter particles can already be correlated at the initial time, we were able to predict $\tau \approx 887.51$ s, which, for a toy model, is impressively close to the values found by real experiments. 

Since the confining volume $V$ in the calculations was essentially a free parameter that, together with the boundary conditions provided by the $\tau$SPECT trap walls, was set by a real experiment, we interpret our results as supporting evidence for the validity of our proposal. In addition, there are also hints from the existing literature to aid our proposal. Ref.~\cite{He:2021wmq} has recently suggested that there is a neutron Purcell effect, i.e, an effect that is also volume dependent, that could be measured in modern experiments. Furthermore, Ref.~\cite{Mampe:1989xx} has used a neutron trap with a variable length and found a lifetime that increases with the mean free path of the neutron. Our result in Eq.~(\ref{eq:FinalResult}) also indicates that $\tau$ grows with the spatial dimensions of a trap.

In order to transform the toy model used in this article into a viable tool for making predictions for experiments and to accurately verify our proposal of a volume dependence of neutron lifetime measurements, there are some improvements required, which can be done in a future more extensive work. Obviously, for a more realistic model, a neutron and its decay products should be described as fermions, i.e., by spinors. Consequently, the formalism from Ref.~\cite{Kading:2022hhc} must be made applicable to fermions. Neutrons and protons can then either be treated as single particles or as composite particles at the quark level. Extending the formalism from Ref.~\cite{Kading:2022hhc} to gauge bosons, this will allow us to move away from the fermion contact interaction and to instead also consider the W bosons emitted by the neutrons during decay. Furthermore, in many places throughout our computations, we have made rough approximations of sums and integrals by using resting neutrons and protons, and by only extracting the terms corresponding to energy conservation. For accurate predictions, we will have to properly evaluate these sums and integrals, such that we will take into account the full permitted momentum spectrum. In addition, we will need to have knowledge of the initial states of the neutrons entering the experiment, including correlations between a neutron and its decay products as we have used in Sec.~\ref{sec:initialcorrel}. When discussing experiments other than $\tau$SPECT, i.e., those with different confinement structures, we must take into account that there will often be no perfect neutron reflections at the confinement walls. This changes the boundary conditions or must be incorporated by computing density matrices at intermediate times after reflections with Ref.~\cite{Kading:2022hhc}. Besides, as we have stated before in Sec.~\ref{sec:DME12}, there are established finite volume effects like a volume dependence of masses. Taking these into account, would lead to additional differences between the density matrix elements and probabilities in infinite and finite volumes. In this context, we can also consider that we are actually working within finite time intervals when doing our computations. While this is usually ignored, see, for example, Ref.~\cite{Peterken:2024ktm}, such confinements to finite times might not be negligible in future discussions and should, in any case, be studied.
Finally, future computations should also predict the neutron lifetime to higher orders in the coupling constant and take into account loop corrections. Though, this will likely require using the time dependent renormalization formalism that was initially discussed in Ref.~\cite{Burrage:2018pyg}.


\begin{acknowledgments}
The author is grateful to H.~Abele, A.~L.~Báez-Camargo, C.~Briddon, D.~Hartley, P.~Millington and M.~Pitschmann for useful discussions.
This research was funded in whole or in part by the Austrian Science Fund (FWF) [10.55776/PAT8564023]. For open access purposes, the author has applied a CC BY public copyright license to any author accepted manuscript version arising from this submission.
\end{acknowledgments}


\appendix


\section{Directly computing density matrices in infinite volumes}
\label{app:SK}

Here, we will provide a more detailed explanation on how Ref.~\cite{Kading:2022hhc} has obtained the result in Eq.~(\ref{eq:oldresult}). In the same manner as described in what follows, the expressions in Eqs.~(\ref{eq:12Scatt}) - (\ref{eq:NeutFirstresutlinfi}) and Eq.~(\ref{eq:densCorrComp}) have also been derived.

Starting from the quantum Liouville equation for a density operator,
\begin{eqnarray}
\label{Eqn:Liouville}
\frac{\partial}{\partial t} \hat{\rho}(t) &=& -\mathrm{i}[\hat{H}(t),\hat{\rho}(t)]
\end{eqnarray}
with solution
\begin{eqnarray}
\label{eqn:SPS2}
\hat{\rho}(t) &=& (\mathrm{T} e^{-\mathrm{i}\int^t_0 dt' \hat{H}(t')})\hat{\rho}(0)(\tilde{\mathrm{T}} e^{\mathrm{i}\int^t_0 dt' \hat{H}(t')})~, 
\end{eqnarray}
Ref.~\cite{Kading:2022hhc} finds the following expression by using the TFD formalism that we have briefly described in Sec.~\ref{sec:DME12}:
\begin{eqnarray}\label{eq:SolutionTFD}
\hat{\rho}^+(t)\kket{1} &=& \text{T} \exp\left\{{-\mathrm{i}\int\limits_{0}^t\widehat{H}(t')dt'}\right\}\hat{\rho}^+(0)\kket{1}~,
\end{eqnarray}
where $\widehat{H}(t) := \hat{H}(t) \otimes \mathbb{I} - \mathbb{I} \otimes \hat{H}(t)$ describes all interactions within the closed system. If we want to compute particular density matrix elements, we can simply use Eq.~(\ref{eq:SolutionTFD}) in order to project the density operator into the basis of interest. For example, in Eq.~(\ref{eq:oldresult}) we need the element $\rho^\infty_{0,2;0,2}(;\mathbf{p}, \mathbf{k}|;\mathbf{p}', \mathbf{k}'|t)$, which means that we have to compute
\begin{eqnarray}
    \rho^\infty_{0,2;0,2}(;\mathbf{p}, \mathbf{k}|;\mathbf{p}', \mathbf{k}'|t) &=& \text{Tr} \ket{;\mathbf{p}', \mathbf{k}';t}\bra{;\mathbf{p}, \mathbf{k};t} \hat{\rho}(t)~.
\end{eqnarray}
In TFD, we can rewrite this trace as
\begin{eqnarray}
\label{eq:TFDtracetaken}
    \rho^\infty_{0,2;0,2}(;\mathbf{p}, \mathbf{k}|;\mathbf{p}', \mathbf{k}'|t) &=& \bbra{1} (\ket{;\mathbf{p}', \mathbf{k}';t}\bra{;\mathbf{p}, \mathbf{k};t} \otimes \hat{\mathbb{I}}) \hat{\rho}^+(t)\kket{1}~.
\end{eqnarray}
Next, we can substitute Eq.~(\ref{eq:SolutionTFD}) into Eq.~(\ref{eq:TFDtracetaken}), replace the Hamiltonian by an action operator corresponding to the action given in Eq.~(\ref{eq:Action12}), and expand the resulting expression up to second order in $\alpha$. In this way, we find
\begin{eqnarray}
\label{eq:DensApp}
\rho^\infty_{0,2;0,2}(;\mathbf{p}, \mathbf{k}|;\mathbf{p}', \mathbf{k}'|t)
&\approx&
- \frac{\alpha^2}{8}\mathcal{M}^2
\langle\langle ;\mathbf{p}_+,\mathbf{k}_+, \mathbf{p}'_-,\mathbf{k}'_-;t|
 \sum\limits_{a,b=\pm} ab \int_{zz'} \hat{\phi}_z^a\hat{\phi}_{z'}^b(\hat{\varphi}_z^a)^2(\hat{\varphi}_{z'}^b)^2
\hat{\rho}^+(0) |1\rangle\rangle~.~~~~~~
\end{eqnarray}
Note that that the right-hand sides of Eqs.~(\ref{eq:DensApp}) and (\ref{eq:Startdisc12}) coincide, which means that, at this point, there is no obvious difference in the computations for finite or infinite volumes. Since we are interested in describing the decay of a single particle $\phi$ into two copies of $\varphi$, we say that, at the initial time $0$, the only non-vanishing density matrix element was $\rho^\infty_{1,0;1,0}(\mathbf{q};|\mathbf{q}';|0)$, such that
\begin{eqnarray}
\rho^\infty_{0,2;0,2}(;\mathbf{p}, \mathbf{k}|;\mathbf{p}', \mathbf{k}'|t)
&\approx&
- \frac{\alpha^2}{8}\mathcal{M}^2
\langle\langle ;\mathbf{p}_+,\mathbf{k}_+, \mathbf{p}'_-,\mathbf{k}'_-;t|
 \sum\limits_{a,b=\pm} ab \int_{zz'} \hat{\phi}_z^a\hat{\phi}_{z'}^b(\hat{\varphi}_z^a)^2(\hat{\varphi}_{z'}^b)^2
 \nonumber
 \\
 &&
 ~~~~~~~~~~~~~~
 \times
 \int \Pi^\phi_q \Pi^\phi_{q'}
\rho^\infty_{1,0;1,0}(\mathbf{q};|\mathbf{q}';|0) |\mathbf{q}_+,\mathbf{q}'_-;;0\rangle\rangle~.~~~~~~
\end{eqnarray}
Note that now there is a clear difference to the finite volume case since, here, we are working with a continuous momentum spectrum. Next, we pull out all creation and annihilation operators, such that we are left with a transition amplitude between two TFD vacuum states, i.e., an expression of the form
\begin{eqnarray}
\rho^\infty_{0,2;0,2}(;\mathbf{p}, \mathbf{k}|;\mathbf{p}', \mathbf{k}'|t)
&\approx&
- \frac{\alpha^2}{8}\mathcal{M}^2
\langle\langle 0| \hat{b}^+_\mathbf{p}(t)\hat{b}^+_\mathbf{k}(t) \hat{b}^-_{\mathbf{p}'}(t)\hat{b}^-_{\mathbf{k}'}(t)
 \sum\limits_{a,b=\pm} ab \int_{zz'} \hat{\phi}_z^a\hat{\phi}_{z'}^b(\hat{\varphi}_z^a)^2(\hat{\varphi}_{z'}^b)^2
 \nonumber
 \\
 &&
 ~~~~~~~~~~~~~~
 \times
 \int \Pi^\phi_q \Pi^\phi_{q'}
\rho^\infty_{1,0;1,0}(\mathbf{q};|\mathbf{q}';|0) 
\hat{a}^{+\dagger}_\mathbf{q}(0)\hat{a}^{-\dagger}_{\mathbf{q}'}(0)
|0\rangle\rangle~,~~~~~~
\end{eqnarray}
where the vacuum state is defined as $\kket{0}:= \kket{;;0}$, and the operators $\hat{a}^{\dagger}$ and $\hat{b}$ are associated to $\phi$ and $\varphi$, respectively. The creators and annihilators can be rewritten in terms of field operators, i.e., for $\phi$ we have 
\begin{eqnarray}
\hat{a}^+_{\mathbf{p}}(t) &=& +\mathrm{i}\int_{\mathbf{x}} e^{-\mathrm{i}\mathbf{p}\cdot\mathbf{x}}\partial_{t,E^\phi_{\mathbf{p}}}\hat{\phi}^+(t,\mathbf{x})~,~~~~
\hat{a}^{+\dagger}_{\mathbf{p}}(t) = -\mathrm{i}\int_{\mathbf{x}} e^{+\mathrm{i}\mathbf{p}\cdot\mathbf{x}} \partial_{t,E^\phi_{\mathbf{p}}}^*\hat{\phi}^+(t,\mathbf{x})~,
\nonumber
\\
\hat{a}^-_{\mathbf{p}}(t) &=& -\mathrm{i}\int_{\mathbf{x}} e^{+\mathrm{i}\mathbf{p}\cdot\mathbf{x}}\partial_{t,E^\phi_{\mathbf{p}}}^*\hat{\phi}^-(t,\mathbf{x})~,~~~
\hat{a}^{-\dag}_{\mathbf{p}}(t) = +\mathrm{i}\int_{\mathbf{x}} e^{-\mathrm{i}\mathbf{p}\cdot\mathbf{x}}\partial_{t,E^\phi_{\mathbf{p}}}\hat{\phi}^{-}(t,\mathbf{x})~,
\end{eqnarray}
where $\partial_{t,E^\phi_{\mathbf{p}}} := \overset{\rightarrow}{\partial}_t - \mathrm{i}E^\phi_{\mathbf{p}}$, and the same for the creation and annihilation operators of $\varphi$. Subsequently, the resulting N-point function can then be translated into path integrals over both branches of the Schwinger-Keldysh closed time path since TFD represents an algebraic version of the Schwinger-Keldysh formalism, such that we obtain
\begin{eqnarray}
\rho^\infty_{0,2;0,2}(;\mathbf{p}, \mathbf{k}|;\mathbf{p}', \mathbf{k}'|t)
&\approx&  
- \frac{\alpha^2}{8}\mathcal{M}^2
\lim_{\substack{x^{0(')}_{(1),(2)}\,\to\, t^{+}\\y^{0(')}\,\to\, 0^-}}
\int d\Pi_{\mathbf{q}}d\Pi_{\mathbf{q}'}\rho^\infty_{1,0;1,0}(\mathbf{q};|\mathbf{q}';|0)
\nonumber
\\
&&\,\,\,\,\,\,
\times
\int_{\mathbf{x}_{(1)}\mathbf{x}'_{(1)}\mathbf{x}_{(2)}\mathbf{x}'_{(2)}\mathbf{y}\mathbf{y}'} 
e^{-\mathrm{i}(\mathbf{p}\cdot\mathbf{x}_{(1)} + \mathbf{k}\cdot\mathbf{x}_{(2)} - \mathbf{p}'\cdot\mathbf{x}'_{(1)} - \mathbf{k}'\cdot\mathbf{x}'_{(2)})+\mathrm{i}(\mathbf{q}\cdot\mathbf{y}-\mathbf{q}'\cdot\mathbf{y}')}
\nonumber
\\
&&\,\,\,\,\,\,
\times
\partial_{x_{(1)}^0,E^\varphi_{\mathbf{p}}}\partial_{x_{(1)}^{0'},E^\varphi_{\mathbf{p}'}}^*\partial_{x^0_{(2)},E^\varphi_{\mathbf{k}}}\partial_{x^{0'}_{(2)},E^\varphi_{\mathbf{k}'}}^*\partial_{y^0,E^\phi_{\mathbf{q}}}^*
\partial_{y^{0'},E^\phi_{\mathbf{q}'}}
\nonumber
\\
&&\,\,\,\,\,\,
\times
\int\mathcal{D}\phi^{\pm}\mathcal{D}\varphi^{\pm} e^{\mathrm{i}\widehat{S}_{\phi}[\phi]+\mathrm{i}\widehat{S}_{\varphi}[\varphi]}\varphi^+_{x_{(1)}}\varphi^-_{x'_{(1)}}\varphi^+_{x_{(2)}}\varphi^-_{x'_{(2)}}
\nonumber
\\
&&\,\,\,\,\,\,
\times
 \sum\limits_{a,b=\pm} ab \int_{zz'} \hat{\phi}_z^a\hat{\phi}_{z'}^b(\hat{\varphi}_z^a)^2(\hat{\varphi}_{z'}^b)^2
 \phi^+_{y}\phi^-_{y'}~,
\end{eqnarray}
where we have introduced the limits for the time coordinates in order to preserve the correct time-ordering, i.e., $x^{0(')}_{(1),(2)}$ approach $t$ from above and $y^{0(')}$ approach $0$ from below. Next, we can use Wick's theorem \cite{Wick} in order to evaluate the path integrals. While doing so, as we have already stated below Eq.~(\ref{eq:12Scatt}), we are only allowed to contract fields with each other that belong to the same branch of the Schwinger-Keldysh closed time path. Consequently, we will only obtain Feynman propagators or their complex conjugates, which are called Dyson propagators. All terms but the one corresponding to the diagram in Fig.~\ref{fig:1to2} will give nonphysical disconnected diagrams, which is why we will not consider them in what follows. Keeping only the physical term, we arrive at
\begin{eqnarray}\label{eq:12Prop}
\rho^\infty_{0,2;0,2}(;\mathbf{p}, \mathbf{k}|;\mathbf{p}', \mathbf{k}'|t)
&\approx&
\alpha^2\mathcal{M}^2\lim_{\substack{x^{0(')}_{(1),(2)}\,\to\, t^{+}\\y^{0(')}\,\to\, 0^-}}
\int d\Pi_{\mathbf{q}}d\Pi_{\mathbf{q}'}\rho^\infty_{1,0;1,0}(\mathbf{q};|\mathbf{q}';|0)
\nonumber
\\
&&\,\,\,\,\,\,
\times
\int_{\mathbf{x}_{(1)}\mathbf{x}'_{(1)}\mathbf{x}_{(2)}\mathbf{x}'_{(2)}\mathbf{y}\mathbf{y}'} 
e^{-\mathrm{i}(\mathbf{p}\cdot\mathbf{x}_{(1)} + \mathbf{k}\cdot\mathbf{x}_{(2)} - \mathbf{p}'\cdot\mathbf{x}'_{(1)} - \mathbf{k}'\cdot\mathbf{x}'_{(2)})+\mathrm{i}(\mathbf{q}\cdot\mathbf{y}-\mathbf{q}'\cdot\mathbf{y}')}
\nonumber
\\
&&\,\,\,\,\,\,
\times
\partial_{x_{(1)}^0,E^\varphi_{\mathbf{p}}}\partial_{x_{(1)}^{0'},E^\varphi_{\mathbf{p}'}}^*\partial_{x^0_{(2)},E^\varphi_{\mathbf{k}}}\partial_{x^{0'}_{(2)},E^\varphi_{\mathbf{k}'}}^*\partial_{y^0,E^\phi_{\mathbf{q}}}^*
\partial_{y^{0'},E^\phi_{\mathbf{q}'}}
\nonumber
\\
&&\,\,\,\,\,\,
\times
\int_{zz'} D_{zy}^\mathrm{F} D_{z'y'}^\mathrm{D}  \Delta^{\rm F}_{x_{(1)}z}\Delta^{\rm F}_{x_{(2)}z}  
\Delta^{\rm D}_{x_{(1)}'z'}\Delta^{\rm D}_{x_{(2)}'z'}  
~,
\end{eqnarray}
where $D$ is a $\phi$-propagator, $\Delta$ is a $\varphi$-propagator, and F and D label Feynman and Dyson propagators, respectively. After substituting explicit expressions for the propagators into Eq.~(\ref{eq:12Prop}), i.e.,
\begin{eqnarray}
 D^\mathrm{F}_{xy} &=& - \mathrm{i}\int \frac{d^4k}{(2\pi)^4} \frac{e^{\mathrm{i}k\cdot (x-y)}}{k^2+M^2-\mathrm{i}\epsilon}
 ~,~~~
 D^\mathrm{D}_{xy} = + \mathrm{i}\int \frac{d^4k}{(2\pi)^4} \frac{e^{\mathrm{i}k\cdot (x-y)}}{k^2+M^2+\mathrm{i}\epsilon}\,\,\,,
\\
 \Delta^\mathrm{F}_{xy} &=& - \mathrm{i}\int \frac{d^4k}{(2\pi)^4} \frac{e^{\mathrm{i}k\cdot (x-y)}}{k^2+m_\varphi^2-\mathrm{i}\epsilon}
 ~,~~~
\Delta^\mathrm{D}_{xy} = + \mathrm{i}\int \frac{d^4k}{(2\pi)^4} \frac{e^{\mathrm{i}k\cdot (x-y)}}{k^2+m_\varphi^2+\mathrm{i}\epsilon}
~,
\end{eqnarray}
all 3-momentum integrals in this equation can be evaluated by exploiting that the integrals over spatial coordinates give
\begin{eqnarray}
\label{eq:AppXint}
    \int_\mathbf{x} e^{\mathrm{i}(\mathbf{p}-\mathbf{k})\mathbf{x} } &=& (2\pi)^3 \delta^{(3)}(\mathbf{p}-\mathbf{k})~,
\end{eqnarray}
while the remaining integrals over the zeroth components of the 4-momenta can be dealt with by using Cauchy's integral formula; see App.~D in Ref.~\cite{Kading:2019vyb} for a step-by-step manual for doing such computations. As a result, we obtain Eq.~(\ref{eq:oldresult}).

\bibliography{Bib}
\bibliographystyle{JHEP}

\end{document}